\documentclass[aps,twocolumn,nofootinbib,superscriptaddress,amsmath,amssymb]{revtex4-2}
\usepackage{graphicx}  
\usepackage{color}
\usepackage{times}
\usepackage{float}
\usepackage{afterpage}
\usepackage[utf8]{inputenc}
\usepackage{bm}
\usepackage{ulem}
\usepackage{multirow}
\usepackage{makecell}
\usepackage{url}
\usepackage{natbib}
\usepackage{mathrsfs}
\usepackage{physics}
\usepackage{comment}
\usepackage[table,xcdraw]{xcolor}
\usepackage{booktabs}
\usepackage{pifont}
\usepackage[colorlinks=true,citecolor=blue,urlcolor=blue,linkcolor=blue]{hyperref}
\usepackage[none]{hyphenat}
\usepackage{microtype}
\usepackage{enumitem}
\usepackage{tikz}
\usetikzlibrary{matrix}
\usetikzlibrary{arrows.meta, positioning, shapes.geometric, calc}
\newcommand{\Lie}{\pounds}

\begin{document}

% Title and author information
\title{Non-radial pulsations of gravitationally coupled two-fluid neutron stars in general relativity}

% Authors and affiliations
\author{Ankit Kumar}
\email{ankitlatiyan25@gmail.com; ankitkumar2@iisc.ac.in}
\affiliation{Department of Physics, Indian Institute of Science, Bangalore 560012, India}
\affiliation{CEICO, Institute of Physics of the Czech Academy of Sciences (FZU), Na Slovance 1999/2, 182 00 Prague 8, Czech Republic}

\author{Daniel A. Caballero}
\email{dac8@illinois.edu}
\affiliation{Illinois Center for Advanced Studies of the Universe \& Department of Physics, \\ University of Illinois at Urbana-Champaign, Urbana, Illinois 61801, USA.}

\author{Hajime Sotani}
\email{sotani@yukawa.kyoto-u.ac.jp}
\affiliation{Department of Mathematics and Physics, Kochi University, Kochi, 780-8520, Japan}
\affiliation{RIKEN Center for Interdisciplinary Theoretical and Mathematical Sciences (iTHEMS), RIKEN, Wako 351-0198, Japan}
\affiliation{Theoretical Astrophysics, IAAT, University of T\"{u}bingen, 72076 T\"{u}bingen, Germany}

\author{Nicol\'{a}s Yunes}
\email{nyunes@illinois.edu}
\affiliation{Illinois Center for Advanced Studies of the Universe \& Department of Physics, \\ University of Illinois at Urbana-Champaign, Urbana, Illinois 61801, USA.}

\date{\today}
%%%%%%%%%%%%%%%%%%%%%%%%%%%%%
% Abstract
\begin{abstract}
Non-radial oscillations of neutron stars provide a powerful probe of stellar structure and relativistic gravity, but a fully general relativistic treatment for gravitationally coupled two-fluid stars with independently conserved currents has so far been lacking. In this work, we develop a fully relativistic framework for polar perturbations of gravitationally coupled two-fluid neutron stars, assuming that the two fluids interact only through the common spacetime and are not coupled by entrainment or direct microphysical interactions. We derive the coupled linear perturbation equations governing the metric and both fluid components, and complete the formulation by establishing the regularity, surface, and exterior matching conditions required for a well-posed oscillation eigenvalue problem. We then implement the resulting system numerically and compute representative polar mode spectra for gravitationally coupled two-fluid stellar models. This implementation provides a practical way to address mode identification in gravitationally coupled two-fluid stars, allowing the fundamental ($\mathsf{f}$) and pressure ($\mathsf{p}$) mode branches of the spectrum to be classified according to their dominant inner- or outer-fluid character through the associated eigenfunctions and their node structure. The formalism developed here provides a foundation for extending relativistic asteroseismology to multi-fluid compact stars and for exploring their potential gravitational-wave signatures in a fully general relativistic setting. \\ 
\end{abstract}
%%%%%%%%%%%%%%%%%%%%%%%%%%%%%

% PACS numbers
%\pacs{Insert PACS numbers here}

\maketitle

% Introduction
%%%%%%%%%%%%%%%%%%%%%%%%%%%%%
\section{Introduction}
\label{sec:1}
%%%%%%%%%%%%%%%%%%%%%%%%%%%%%
Neutron stars are now firmly established as sources of gravitational waves across a wide range of dynamical scenarios, extending well beyond the inspiral phase of compact binaries. Whenever the stellar interior or the surrounding spacetime is driven out of equilibrium—by violent events such as gravitational collapse, the formation and relaxation of post-merger remnants, magnetar flares, crustal failures, or sudden angular-momentum rearrangements—the star responds through global stellar oscillations that can imprint characteristic spectral features on the gravitational-wave emission. Unlike chirp signals, which primarily reflect orbital dynamics, oscillation-driven radiation is directly linked to the internal structure of the star and therefore provides a potential observational channel for probing dense-matter physics through mode frequencies and damping times~\cite{PhysRevLett.77.4134, BFSchutz1989}. Although the detectability of such signals depends sensitively on excitation efficiency, dissipation mechanisms, and instrumental sensitivity—particularly in the kilohertz regime—the increasing precision of gravitational-wave interferometers and the development of dedicated analysis strategies have brought stellar asteroseismology into closer contact with observational science. In this context, a reliable theoretical description of neutron star oscillations is essential for establishing a robust connection between prospective gravitational-wave data and the microphysical properties of matter at supranuclear densities.

Non-radial oscillations of relativistic stars provide a direct window into the dynamical coupling between matter and spacetime, encoding detailed information about the internal structure and composition of neutron stars. In general relativity, these oscillations decompose into axial (odd-parity) and polar (even-parity) sectors, distinguished by their transformation properties under parity. The polar sector describes genuine pulsational degrees of freedom of the fluid and couples self-consistently to gravitational radiation, giving rise to families of modes, such as the fundamental ($\mathsf{f}$), pressure ($\mathsf{p}$), gravity ($\mathsf{g}$), and spacetime ($\mathsf{w}$) modes~\cite{10.1093/mnras/101.8.367, Kokkotas1999, 1983ApJS...53...73L}. These polar oscillations are of particular interest for relativistic stellar asteroseismology, as they dominate the star’s gravitational-wave response to dynamical perturbations and carry imprints of the underlying microphysics through their characteristic frequencies and damping properties~\cite{1967ApJ...149..591T}. Axial perturbations, by contrast, do not involve compressional motion of the fluid. In non-rotating stars, the axial fluid degrees of freedom correspond only to stationary, toroidal motions without a restoring force, while the associated gravitational-wave solutions describe spacetime oscillations with only very weak coupling to the stellar fluid~\cite{1970ApJ...159..847C, 1994MNRAS.268.1015K}. For this reason, polar non-radial oscillations constitute the physically relevant channel for studying neutron-star dynamics and gravitational-wave emission~\cite{1969ApJ...155..163P, PhysRevLett.77.4134}, and form the focus of the present work.

The theoretical description of polar non-radial oscillations in single-fluid relativistic stars is by now well established and has been developed systematically within general relativity~\cite{1970ApJ...159..847C, 1967ApJ...149..591T, 1969ApJ...155..163P, 1983ApJS...53...73L, 1985ApJ...292...12D}. In this framework, the coupled linear perturbations of the spacetime metric and the stellar fluid can be formulated as a closed system of linearized Einstein–matter equations, yielding a discrete spectrum of oscillation modes with well-defined physical interpretation. Extensive studies have clarified the properties of the fluid-dominated mode families (including the $\mathsf{f}$, $\mathsf{p}$, and, when stratification is present, $\mathsf{g}$-modes) as well as spacetime-led $\mathsf{w}$-modes; these studies have demonstrated how the modes' frequencies and damping times depend sensitively on the star's equation of state, stellar compactness, and internal composition~\cite{10.1046/j.1365-8711.1998.01840.x, 10.1093/mnras/222.3.393, 1983ApJS...53...73L, 10.1093/mnras/stx3067, 10.1046/j.1365-8711.1999.02983.x, PhysRevD.70.124015, Jaikumar:2021jbw, PhysRevD.83.024014, PhysRevD.80.023011, PhysRevC.106.015805}. The fundamental ($\mathsf{f}$--) mode corresponds to a global oscillation of the stellar fluid coupled strongly to spacetime curvature, while the pressure ($\mathsf{p}$--) modes are predominantly governed by the fluid’s compressibility and sound speed. Gravity ($\mathsf{g}$--) modes arise when stable stratification or composition gradients are present~\cite{10.1093/mnras/227.2.265, PhysRevD.65.024010}, reflecting buoyancy-driven restoring forces, whereas spacetime ($\mathsf{w}$--) modes are dominated by oscillations of the gravitational field itself with only weak participation of the fluid~\cite{10.1093/mnras/255.1.119, PhysRevD.48.3467}. For completeness, rotating neutron stars also support axial-led inertial oscillations,  such as $\mathsf{r}$--modes, which belong to the odd-parity sector and are restored primarily by the Coriolis force~\cite{1998ApJ...502..708A, Kojima_1999}. Together, these mode families provide complementary probes of neutron-star structure and have made non-radial asteroseismology a powerful theoretical tool for connecting microscopic physics to observable spectral properties. This standard construction, however, is rooted in a physical description where the stellar interior is treated as a single effective fluid, characterized by a unique four-velocity and thermodynamic response. In systems where multiple fluids coexist and evolve dynamically within the same gravitational potential, this description must be generalized, and the structure of the non-radial perturbation problem acquires qualitatively new features.

Neutron star interiors are not generically restricted to a single comoving fluid, and multi-fluid dynamics arises naturally in several physically-motivated contexts. A well-known example is provided by superfluid neutron stars~\cite{1988ApJ...333..880E, BAYM1969, Lombardo2001}, where superfluid neutrons can move relative to a charged component, leading to a genuine multi-velocity description of the stellar interior. Relativistic multi-fluid frameworks that explicitly incorporate entrainment have been developed to describe such systems, and non-radial oscillations of superfluid stars have been formulated within these approaches~\cite{10.1111/j.1365-2966.2008.13426.x, 1994ApJ...421..689L}. In these theories, the dynamics is derived from a variational principle built on a master function of the scalar invariants constructed from the constituent currents, with entrainment coefficients encoding the microphysical coupling between them and determining the structure of the perturbation equations and mode spectrum~\cite{PhysRevD.60.104025, PhysRevD.66.104002, PhysRevD.84.107301, PhysRevD.78.083008}. However, there also exist physically well-defined scenarios in which multiple components coexist within a neutron star while remaining dynamically independent at the microphysical level, interacting only through the common gravitational field. Dark matter admixed neutron stars provide a particularly well-defined realization of this regime, as the dark component possesses its own conserved current and four-velocity while coupling to ordinary matter predominantly via gravity~\cite{PhysRevD.111.083038, GOLDMAN2013200, Kain:2021hpk, Shawqi_2024, SANDIN2009278, PhysRevD.107.115028}. In such systems, the stellar dynamics cannot be reduced to an effective single-fluid description, and it is not generally captured by standard entrainment-based superfluid formalisms by merely setting entrainment terms to zero, because the physical content and typical assumptions differ (e.g., distinct constituents and, in general, distinct radial supports). This reflects fundamental differences in the underlying variational structure, the nature of the conserved currents, and the physical interpretation of counter-moving degrees of freedom, which persist even in the absence of direct microphysical interactions and are further accentuated in configurations where the two components occupy distinct radial regions. As a result, the presence of two independently evolving fluids introduces additional dynamical degrees of freedom that must be incorporated self-consistently into the description of non-radial oscillations and are expected to qualitatively modify the structure of the mode spectrum.

Despite substantial progress in related directions, a specific and physically important problem remains unaddressed. Non-radial oscillations of neutron stars have been extensively studied in the single-fluid framework, and relativistic multi-fluid formulations with entrainment have enabled detailed investigations of superfluid oscillation spectra~\cite{1988ApJ...325..725M, refId0, Stavridis:2007xz, PhysRevD.70.124015, Pratten2020, PhysRevD.101.103009, PhysRevD.103.123015, PhysRevD.106.123002, Doneva:2013zqa, PhysRevC.108.015803, Sotani:2012qc, 10.1111/j.1365-2966.2009.14734.x, 10.1111/j.1365-2966.2008.13426.x, 10.1111/j.1365-2966.2011.19725.x, Samuelsson_2009}. In parallel, gravitationally coupled two-fluid neutron stars—particularly in the context of dark matter admixture—have been explored primarily at the level of equilibrium structure, radial oscillations, and stability analyses~\cite{Kumar:2025yei, Caballero:2024qtv, Kumar:2025oyx, Gleason:2022eeg, PhysRevD.111.123034, PhysRevD.97.123007, Sotani:2025lzy}. Non-radial oscillations of dark matter admixed neutron stars have also been investigated in the Cowling approximation~\cite{Thakur_2024}, providing an important step toward using non-radial mode spectra as probes of dark matter. In that approach, however, although the equilibrium configuration is constructed with two fluid components, the perturbation sector is not formulated as a fully coupled two-fluid problem, with independent displacement fields for both fluids. A consistent two-fluid Cowling treatment, in which the dynamical degrees of freedom of both components are retained, was developed later in~\cite{Sotani:2025hzb}. 

To the best of our knowledge, a fully relativistic formulation of non-radial perturbations for gravitationally coupled two-fluid stars, in which the constituents interact solely through the spacetime geometry and possess independently conserved currents, has not yet been systematically developed. The absence of such a framework limits our ability to analyze the dynamical response of multi-component stars beyond radial motion and to determine how the presence of multiple fluid degrees of freedom in the stellar interior modifies the structure of the non-radial oscillation spectrum. In the presence of multiple independently evolving fluids, the perturbation equations involve multiple displacement fields and metric perturbations, and the associated eigenvalue problem is structurally distinct from its single-fluid counterpart. Addressing this gap is essential for extending relativistic asteroseismology to physically well-motivated two-fluid systems and for enabling consistent comparisons between theoretical predictions and potential gravitational-wave signatures from multi-component neutron stars.

In this work, we develop a fully relativistic formulation of polar non-radial perturbations for gravitationally coupled, two-fluid neutron stars. Starting from a two-fluid system with independently conserved currents, we derive the complete set of linearized perturbation equations governing the coupled dynamics of the spacetime metric and both fluid components in a fully general relativistic framework, without invoking the Cowling approximation~\cite{1983ApJ...268..837M}. The analysis is restricted to the even-parity sector and assumes that the two fluids interact exclusively through the spacetime geometry, with no entrainment or direct microphysical coupling between them. The resulting formalism yields a closed, self-consistent system of equations suitable for studying non-radial oscillations in a broad class of two-fluid configurations. As an application of the framework, we implement the perturbation equations numerically and compute representative polar mode spectra for gravitationally coupled, two-fluid neutron star models, demonstrating how the formalism can be employed in practice to analyze non-radial stellar oscillations, and extending earlier studies~\cite{Sotani:2025hzb} performed within the Cowling approximation to a fully relativistic treatment. 
The representative mirror dark matter models considered in this work show that the additional fluid degree of freedom leaves a measurable imprint on the polar spectrum, producing distinct mode branches associated with the two fluid components. In particular, the ordinary-matter-led $\mathsf{f}$--mode branch is shifted by the gravitational presence of the dark component, and the equation of state-insensitive mass-scaled $\mathsf{f}$--mode universal relation with compactness, known for standard single-fluid neutron stars, is not preserved in the same form for gravitationally coupled two-fluid configurations.

The paper is organized as follows. In Section~\ref{sec:2}, we formulate the fully relativistic theory of polar non-radial perturbations for gravitationally coupled two-fluid neutron stars. We begin by specifying the static, spherically symmetric background configuration of independently conserved fluids interacting solely through the spacetime geometry (Sec.~\ref{sec:2a}), and then derive the coupled linear perturbation equations for the metric and both fluid components in the even-parity sector (Sec.~\ref{sec:2c}). The formulation is completed by imposing the appropriate regularity conditions at the center, boundary conditions at the stellar surfaces of each fluid (Sec.~\ref{sec:2d}), and matching conditions to the exterior vacuum spacetime (Sec.~\ref{sec:2e}). In Section~\ref{sec:3}, we describe the numerical implementation of the perturbation formalism and the procedure used to compute the non-radial mode spectrum. In Section~\ref{sec:4}, we apply the framework to mirror dark matter admixed neutron star models, illustrating the practical application of the framework and discussing the resulting mode structure and representative asteroseismic features. We conclude in Section~\ref{sec:5} with a summary and comment on future directions.

Throughout this paper, we will use the following conventions. Greek indices denote spacetime coordinates, and uppercase Latin indices, such as `$X,\ Y = 1, 2, \dots, N$', label the different fluid components. Summation over fluid labels is written explicitly and is not implied by repetition. We adopt the metric signature $(-,+,+,+)$ and work in geometric units with $G = c = 1$. The stellar background is assumed to be static and spherically symmetric, and the matter content is described by multiple perfect fluids with independently conserved currents, interacting solely through the spacetime geometry. No entrainment or direct microphysical coupling between the fluids is included. We restrict attention to polar (even-parity) perturbations, which describe genuine stellar pulsations and couple self-consistently to gravitational radiation.

%%%%%%%%%%%%%%%
%%%%%%%%%%%%%%%%%%%%%%%%%%%%%
\section{Non-radial perturbations of gravitationally coupled two-fluid neutron stars}
\label{sec:2}
%%%%%%%%%%%%%%%%%%%%%%%%%%%%%
%%%%%%%%%%%%%%%

%%%%%%%%%%%%%%%
%%%%%%%%%%%%%%%%%%%%%%%%%%%%%
\subsection{Background two-fluid stellar configuration}
\label{sec:2a}
%%%%%%%%%%%%%%%%%%%%%%%%%%%%%
%%%%%%%%%%%%%%%
We consider a static, spherically symmetric spacetime describing a compact star composed of $N$ perfect-fluid components that interact only through gravity. Each component $X=1, 2, \dots, N$ is characterized by its own energy density ${\cal E}_{X}$, pressure $p_{X}$, and four-velocity $u^\mu_{X}$, and is assumed to possess an independently conserved stress–energy tensor. In equilibrium, the configuration is static, and each fluid is at rest with respect to the chosen coordinate system, so that the background four-velocity of every component is purely timelike and aligned with the timelike Killing vector of the spacetime. Writing the metric in the standard form
%%%%%%%%%%%%%%%
\begin{equation}
    ds^2=-e^{2\Phi(r)} dt^2+e^{2\Lambda(r)}dr^2+r^2(d\theta^{2}+\sin^{2}\theta\, d\phi^{2}) \ ,
\end{equation}
%%%%%%%%%%%%%%%
the background four-velocity of each component is therefore
%%%%%%%%%%%%%%%
\begin{equation}
    u_{X}^{\mu} = e^{-\Phi(r)}\delta_{t}^{\mu}\, , \quad\quad\quad u_{\mu}^{X} u_{X}^{\mu} = -1
\end{equation}
%%%%%%%%%%%%%%%
so that the equilibrium state is fully specified by the radial profiles $\Phi(r)$ and $\Lambda(r)$, and the thermodynamic variables ${\cal E}_{X}$ and $p_{X}$ for each fluid. The stress–energy tensor of each component is taken to be that of a perfect fluid,
%%%%%%%%%%%%%%%
\begin{equation}
    T^{\mu\nu}_{X}=({\cal E}_{X}+p_{X})u^\mu_{X}u^\nu_{X}+p_{X}g^{\mu\nu} \ .
\end{equation}
%%%%%%%%%%%%%%%
We further assume that each component obeys its own barotropic equation of state $p_{X}=p_{X} ({\cal E}_{X})$, with no entrainment or direct microphysical coupling between fluids. Since the interaction is purely gravitational, each component is separately conserved, 
%%%%%%%%%%%%%%%
\begin{equation}
    \nabla_\mu T^{\mu\nu}_{X}=0\ , \quad \quad \quad {\text{for each}}\ X = 1, 2, \dots , N
\end{equation}
%%%%%%%%%%%%%%%
which, upon projection along $u^{\mu}_{X}$ and orthogonally to it, yields, respectively, the energy-conservation equation and the relativistic Euler equation for each component,
%%%%%%%%%%%%%%%
\begin{align}
    &u^\nu_{X}\nabla_\nu\, {\cal E}_{X}+({\cal E}_{X}+p_{X})\nabla_\nu\, u^\nu_{X}=0 \ , \label{eq:energy}\\
    &({\cal E}_{X}+p_{X})u^\nu_{X}\nabla_\nu\, u^\mu_{X}+(g^{\mu\nu}+u^\mu_{X}u^\nu_{X})\nabla_\nu\, p_{X}=0\, . \label{eq:euler}
\end{align}
%%%%%%%%%%%%%%%
Moreover, we introduce a conserved particle-number current $n_{X}u^\mu_{X}$ for each component, where $n_{X}$ denotes the particle number density of fluid $X$. The conservation of particle number is expressed as~\cite{Caballero:2024qtv}
%%%%%%%%%%%%%%%
\begin{equation}
    \nabla_\mu(n_{X} u^\mu_{X})=0 \ ,
\end{equation}
%%%%%%%%%%%%%%%
which provides an independent conservation law for each fluid. For a barotropic component whose energy density can be regarded as a function of the number density, ${\cal E}_{X} = {\cal E}_{X} (n_{X})$, the particle-number conservation law together with the energy-balance equation [Eq.~\eqref{eq:energy}] implies the differential thermodynamic relation
%%%%%%%%%%%%%%%
\begin{equation}
    d{\cal E}_{X} = \frac{{\cal E}_{X}+p_{X}}{n_{X}}dn_{X} \ ,
\end{equation}
%%%%%%%%%%%%%%%
which closes the thermodynamic sector of the background and perturbed equations for each fluid.

The different fluids are coupled exclusively through the spacetime geometry via the Einstein field equations,
%%%%%%%%%%%%%%%
\begin{equation}
    G^{\mu}_{\, \nu}=8\pi \sum_{X} T^{\mu}_{\ \nu\ X} \ .
\end{equation}
%%%%%%%%%%%%%%%
For a static, spherically symmetric configuration, the Einstein equations together with the Euler equations for each component yield the multi-fluid generalization of the Tolman–Oppenheimer–Volkoff (TOV) equations~\cite{GOLDMAN2013200, Kumar:2025yei, Caballero:2024qtv, 559f-kl69},
%%%%%%%%%%%%%%%
\begin{align}
    \frac{d\Phi}{dr}&=\frac{1}{r^2}\frac{1}{1-\tfrac{2m}{r}}\left(m+\sum_{X}4\pi r^3 p_{X}\right) \ , \label{eq:TOVphi}\\
    \frac{dm}{dr}&=4\pi r^2\sum_{X}{\cal E}_{X} \ , \label{eq:TOVm}\\
    \frac{dp_{X}}{dr}&=-({\cal E}_{X}+p_{X})\frac{d\Phi}{dr} \ , \label{eq:TOVp}
\end{align}
%%%%%%%%%%%%%%%
where the mass function $m(r)$ is defined through the metric relation $e^{-2\Lambda(r)} = 1 - 2m/r$.

Together with the equations of state for the individual components, the multi-fluid TOV system defines a well-posed boundary-value problem that can be integrated from the center of the star outward. The integration for each component terminates at a radius $R_{X}$ determined by the condition $p_{X}(R_{X})=0$, beyond which its pressure and energy density vanish identically. The outer boundary of the matter distribution is therefore located at
%%%%%%%%%%%%%%%
\begin{equation}
    R_{\rm{out}} \equiv \max\limits_{X} \{R_{X}\}\ ,
\end{equation}
%%%%%%%%%%%%%%%
and the total gravitational mass of the configuration is given by $M = m(R_{\rm{out}})$. For $r > R_{\rm{out}}$, the spacetime is described by the exterior Schwarzschild solution with mass parameter $M$. This construction fixes the radial domain of the interior problem and specifies the matching surface to the exterior vacuum region that will be used in the formulation of the non-radial perturbation equations.
%%%%%%%%%%%%%%%
%%%%%%%%%%%%%%%%%%%%%%%%%%%%%
\subsection{Perturbation variables and kinematics}
\label{sec:2b}
%%%%%%%%%%%%%%%%%%%%%%%%%%%%%
%%%%%%%%%%%%%%%
We now consider linear perturbations of the spacetime geometry and of each fluid about the static, spherically symmetric background described in Sec.~\ref{sec:2a}. We distinguish between Eulerian perturbations, denoted by $\delta$, and Lagrangian perturbations following the motion of a given fluid $X$, denoted by $\Delta_{X}$. The two are related through the Lagrangian displacement vector $\xi_{X}^{\mu}$, which describes the displacement of fluid elements away from their equilibrium positions under the perturbation. For any tensor quantity $T$ defined on the background spacetime, the Lagrangian and Eulerian variations are related by
%%%%%%%%%%%%%%%
\begin{equation}\label{eq:L&E_pert_rltn}
    \Delta_X T=\delta T + \Lie_{\xi_X} T \ ,
\end{equation}
%%%%%%%%%%%%%%%
where $\Lie_{\xi_X}$ denotes the Lie derivative along $\xi_{X}^{\mu}$. In particular, for a background scalar function $f(r)$, one has $\Delta_{X} f = \delta f + \xi_{X}^{\mu}\nabla_{\mu} f$. 

The Lagrangian displacement $\xi_{X}^{\mu}$ is not unique, since adding any component parallel to the background four-velocity $u_{X}^{\mu}$ corresponds to a relabeling of fluid elements along the unperturbed flow and does not change the physical perturbation. This reflects the invariance of the Lagrangian description under shifts of the displacement along the background fluid flow. We fix this gauge freedom by imposing the orthogonality condition $u^\mu_X \xi_\mu^X = 0$, so that the displacement vector is purely spatial in the local rest frame of fluid $X$. This choice provides a unique and physically transparent definition of the displacement associated with each fluid perturbation.

Using these definitions, together with the normalization $u_{X}^{\mu}u^{X}_{\mu} = -1$, which fixes the perturbation of the four-velocity in terms of the metric perturbation, and the thermodynamic relations of Sec.~\ref{sec:2a}, the Lagrangian perturbations of the fluid variables can be written in terms of the metric perturbations and the displacement fields (see, e.g., Ref.~\cite{Friedman1978}) as 
%%%%%%%%%%%%%%%
\begin{subequations}\label{eq:fluidperts}
\begin{align}
    \frac{\Delta_X n_X}{n_X} &= -\frac{1}{2}(u^\alpha_Xu^\beta_X +g^{\alpha\beta})\Delta_X g_{\alpha\beta} \ , \label{eq:pertn}\\
    \Delta_X p_X &= \gamma_X p_X\frac{\Delta_X n_X}{n_X}\ , \label{eq:pertp}\\
    \Delta_X {\cal E}_X &= ({\cal E}_X+p_X)\frac{\Delta_{X} n_X}{n_X}\ ,\\
    \Delta_X u_X^\mu &= \frac{1}{2}u^\mu_X u^\alpha_X u^\beta_X \Delta_X g_{\alpha\beta} \ .
\end{align}
\end{subequations}
%%%%%%%%%%%%%%%
In Eqs.~\eqref{eq:fluidperts} and in what follows, any quantity that does not carry either an Eulerian variation symbol $\delta$ or a Lagrangian variation symbol $\Delta_{X}$ refers to a \textit{background} (equilibrium) quantity. Here, the adiabatic index of fluid $X$ is defined by 
%%%%%%%%%%%%%%%
\begin{equation}
    \gamma_{X} \equiv \frac{\partial \ln p_{X}}{\partial \ln n_{X}} \ .
\end{equation}
%%%%%%%%%%%%%%%
For a barotropic equation of state, this can be written equivalently in terms of the sound speed as 
%%%%%%%%%%%%%%%
\begin{equation}
    \gamma_{X} = c_{X}^{2} \frac{{\cal E}_{X} + p_{X}}{p_{X}}\ ,
\end{equation}
%%%%%%%%%%%%%%%
where $c_{X}^{2} \equiv \partial p_{X}/ \partial {\cal E}_{X}$ is the squared sound speed of fluid $X$.

The perturbed Euler equation for each fluid is obtained most easily by taking the Eulerian perturbation of Eq.~\eqref{eq:euler} and retaining only terms linear in the perturbations. Introducing the four-acceleration $a_{X}^{\mu} \equiv u^\nu_X\nabla_\nu u^\mu_X$, the Eulerian variation of Eq.~\eqref{eq:euler} yields
%%%%%%%%%%%%%%%
\begin{align}\label{eq:perteul}
    ({\cal E}_X+&p_X) \delta a_{X}^{\mu} +(\delta{\cal E}_X+\delta p_X) a_{X}^{\mu}+(g^{\mu\nu} +u^\mu_X u^\nu_X)\nabla_\nu \delta p_X  \nonumber\\
    &-\left[h_{\alpha\beta}g^{\alpha\mu}g^{\beta\nu} -\delta (u^\mu_X u^\nu_X)\right]\nabla_\nu p_X=0 \ .
\end{align}
%%%%%%%%%%%%%%%
Here $h_{\mu\nu} \equiv \delta g_{\mu\nu}$ denotes the Eulerian metric perturbation, and indices on $h_{\mu\nu}$  are raised/lowered with the background metric.

Together, the relations above provide a complete kinematical framework for describing linear perturbations of the fluid variables in terms of the metric perturbations and the displacement fields of each fluid component. These expressions will be combined with the perturbed Einstein equations in the following subsection, where we specialize to polar perturbations of a static, spherically symmetric background and introduce a harmonic decomposition of both the spacetime and fluid perturbations.
%%%%%%%%%%%%%%%
%%%%%%%%%%%%%%%%%%%%%%%%%%%%%
\subsection{Even-parity (polar) perturbations and perturbation equations}
\label{sec:2c}
%%%%%%%%%%%%%%%%%%%%%%%%%%%%%
%%%%%%%%%%%%%%%
As a consequence of Eq.~\eqref{eq:fluidperts}, the perturbations of the fluid variables can be expressed in terms of the Lagrangian displacement vector $\xi^\mu_X$ and the Eulerian perturbation of the spacetime metric, $h_{\mu\nu} \equiv \delta g_{\mu\nu}$. The remaining freedom in the metric perturbation can be fixed by an appropriate choice of gauge for the problem. In what follows, the perturbations are decomposed into spherical harmonics with a harmonic time dependence, and we restrict attention to even-parity modes with $l\geq 2$. Adopting the Regge–Wheeler gauge for the metric decomposition, the non-vanishing components of the metric perturbation are written as~\cite{1985ApJ...292...12D,1967ApJ...149..591T}
%%%%%%%%%%%%%%%
\begin{subequations}
    \begin{align}
        h_{00}&=\sum_{l,m}e^{2\Phi}r^lH_0^{(lm)} Y^{lm}e^{i\omega t} \ ,\\
        h_{01}&=\sum_{l,m}i\omega r^{l+1}H_1^{(lm)} Y^{lm}e^{i\omega t}=h_{10} \ ,\\
        h_{11}&=\sum_{l,m}e^{2\Lambda}r^lH_2^{(lm)} Y^{lm} e^{i\omega t} \ ,\\
        h_{22}&=\sum_{l,m}r^{l+2} K^{(lm)} Y^{lm} e^{i\omega t}=\csc^2\theta\ h_{33} \ .
    \end{align}
\end{subequations}
%%%%%%%%%%%%%%%
The functions $H_0^{(lm)} (r)$, $H_1^{(lm)} (r)$, $H_2^{(lm)} (r)$ and $K^{(lm)} (r)$ represent the radial amplitudes of the metric perturbation functions.

For the Lagrangian displacement vector, we impose the spatial gauge condition $\xi^\mu_X u_\mu^X=0$ (introduced in Sec.~\ref{sec:2b}), which ensures that the displacement is purely spatial in the rest frame of each fluid. Restricting to the even-parity sector, the displacement vector can then be decomposed into spherical harmonics as
%%%%%%%%%%%%%%%
\begin{subequations}
\begin{align}
    \xi^r_X&=\sum_{l,m} r^{l-1} e^{-\Lambda} W_X^{(lm)} Y^{lm}e^{i\omega t} \ ,\\
    \xi^\theta_X&=-\sum_{l,m}r^{l-2}V_X^{(lm)}\partial_\theta Y^{lm}e^{i\omega t} \ ,\\
    \xi^\varphi_X&=-\sum_{l,m}r^{l-2}\frac{V_X^{(lm)}}{\sin^2\theta}\partial_\varphi Y^{lm}e^{i\omega t} \ .
\end{align}
\end{subequations}
%%%%%%%%%%%%%%%
The fluid perturbations are described by the functions $W_X^{(lm)} (r)$ and $V_X^{(lm)} (r)$, which correspond to the radial and angular components of the Lagrangian displacement of each fluid. In what follows, we suppress the spherical harmonic indices ($l,m$), and all perturbation variables are understood to correspond to a fixed harmonic mode, with the common time dependence $e^{i\omega t}$ implicitly factored out.

Substituting the above decompositions into the linearized Einstein equations yields the coupled equations governing the polar perturbations. The geometric sector, represented by the perturbed Einstein tensor, retains the same form as in the single-fluid problem, since the introduction of additional fluids affects only the matter source through the perturbed stress–energy tensor. Thus, the multi-fluid structure enters the perturbed field equations explicitly through the source term $\delta T^{\mu}_{\, \nu} = \sum_{X} \delta {T^{\mu}_{\, \nu}}_{X}$, while the differential operator acting on the metric perturbations is the same as in the single-fluid case evaluated on the multi-fluid background. As in the single-fluid case, the angular combination $\delta G_{\theta\theta}-\csc^2\theta\ \delta G_{\varphi\varphi}=0$ implies, for $l\geq 2$, that $H_2=H_0$~\cite{1983ApJS...53...73L,1967ApJ...149..591T}. We use this relation throughout the remainder of the derivation and eliminate $H_{2}$ in favour of $H_{0}$. The relevant components of the perturbed Einstein tensor take the form
%%%%%%%%%%%%%%%
\begin{subequations}
    \begin{align}
        \delta G_t^{\ r} = &i\omega r^le^{-2\Lambda}\bigg[\frac{l(l+1)}{2r}H_1+\frac{1}{r}H_0 \nonumber\\
        & \quad\quad\quad\quad-\left(\frac{l+1}{r}-\Phi'\right)K -\frac{dK}{dr}\bigg]Y^{lm} \ ,\\
        \delta G_t^{\ \theta} =& -\frac{i\omega}{2}r^{l-2} e^{-2\Lambda} \bigg\{e^{2\Lambda}K+[(\Lambda'-\Phi')r-(l+1)]H_1 \nonumber\\
        &\quad \quad \quad \quad \quad \quad + e^{2\Lambda}H_0-r\frac{dH_1}{dr}\bigg\}\partial_\theta Y^{lm} \ .
    \end{align}
\end{subequations}
%%%%%%%%%%%%%%%
The corresponding components of the perturbed stress–energy tensor take the form
%%%%%%%%%%%%%%%
\begin{subequations}
    \begin{align}
        \delta T_t^{\ r}&=-i\omega e^{-\Lambda}r^{l-1}\sum_X({\cal E}_X+p_X)W_X Y^{lm}\ ,\\
        \delta T_t^{\ \theta}&=i\omega r^{l-2} \sum_X ({\cal E}_X+p_X) V_X\partial_\theta Y^{lm} \ .       
    \end{align}
\end{subequations}
%%%%%%%%%%%%%%%
Equating these expressions through the linearized Einstein equations, $\delta G^{\mu}_{\, \nu} = 8\pi\ \delta T^{\mu}_{\, \nu}$, yields
%%%%%%%%%%%%%%%
\begin{align}
    \frac{dK}{dr} =& -\left(\frac{l+1}{r}-\Phi'\right)K+\frac{l(l+1)}{2r}H_1+\frac1rH_0 \nonumber\\
    & +\frac{8\pi e^{\Lambda}}{r}\sum_X({\cal E}_X+p_X)W_X \ , \label{eq:dK} \\
    \frac{dH_1}{dr} =&\ H_1\left[(\Lambda'-\Phi')-\frac{l+1}{r}\right]+\frac{e^{2\Lambda}}{r}\left(H_0+K\right) \nonumber\\
    & +\frac{16\pi e^{2\Lambda}}{r}\sum_X({\cal E}_X+p_X)V_X\ . \label{eq:dH1}   
\end{align}
%%%%%%%%%%%%%%%
These two relations, corresponding to the $tr$ and $t\theta$ components of the linearized Einstein equations, determine the metric amplitudes 
$K$ and $H_{1}$ in terms of the fluid displacements and provide the first part of the coupled perturbation system.

With the metric and displacement decompositions specified, the fluid perturbations can now be written explicitly. Substituting the harmonic expansions into Eq. \eqref{eq:pertn}, one obtains
%%%%%%%%%%%%%%%
\begin{align}\label{eq:deln}
    \frac{\Delta_X n_X}{n_X} =& -\left(\frac12 H_0+K\right)-\frac{l(l+1)}{r^2} V_X -e^{-\Lambda}\frac{l+1}{r^2}W_X \nonumber\\
    &\ -e^{-\Lambda}\frac{1}{r}\frac{dW_X}{dr} \ .
\end{align}
%%%%%%%%%%%%%%%
Through the thermodynamic relation in Eq.~\eqref{eq:pertp}, this expression also determines the Lagrangian pressure perturbation of each fluid. It is therefore convenient to define the variable\footnote{Here and in what follows, the harmonic indices are suppressed. More explicitly, $\Delta_X p_X=-e^{-\Phi}\sum_{l,m}r^l\mathcal P_X^{(lm)}Y^{lm}e^{i\omega t}$. A similar suppression is understood in Eq.~\eqref{eq:deln}.} $\mathcal P_X\equiv -\Delta_Xp_Xe^{\Phi}$, since the surface condition for each fluid is simply $\Delta_Xp_X(R_X)=0$. Using Eq. \eqref{eq:pertp}, Eq. \eqref{eq:deln} can then be rewritten in terms of $\mathcal{P}_{X}$, which yields
%%%%%%%%%%%%%%%
\begin{align}\label{eq:dW}
    \frac{dW_X}{dr} =& -\frac{l+1}{r}W_X-\frac{l(l+1)}{r}e^{\Lambda}V_X-re^{\Lambda}\left(\frac{H_{0}}{2}+K\right) \nonumber\\
    &\ +\frac{re^{\Lambda-\Phi}}{\gamma_Xp_X}\mathcal P_X \ .
\end{align}
%%%%%%%%%%%%%%%
The remaining equation for the fluid sector follows from the radial component of the perturbed Euler equation, Eq.~\eqref{eq:perteul}. After substituting the harmonic expansions and eliminating the remaining perturbation variables in favor of $W_{X}$, $V_{X}$, $H_{0}$, $H_{1}$, and $K$, one obtains the following first-order equation\footnote{Intermediate algebraic steps are outlined in Appendix~\ref{app:eqs}.} for ${\mathcal P}_X$:
%%%%%%%%%%%%%%%
\begin{align}\label{eq:dP}
    \frac{d\mathcal P_X}{dr} =& -\frac lr\mathcal P_X+({\cal E}_X+p_X)e^\Phi\Bigg\{\frac12\left(\Phi'-\frac1r\right)H_0 \nonumber \\
    &\ -\frac12\left(\omega^2re^{-2\Phi}+\frac{l(l+1)}{2r}\right)H_1 \nonumber \\
    &\ +\left(\frac{1}{2r}-\frac32\Phi'\right)K -\frac{l(l+1)}{r^2}\Phi'V_X \nonumber\\
    &\ -\frac1r\left[\omega^2e^{\Lambda-2\Phi}-r^2\frac{d}{dr}\left(\frac{1}{r^2}e^{-\Lambda}\Phi'\right)\right]W_X\Bigg\} \nonumber \\
    &\ -\frac{4\pi}{r}e^{\Lambda+\Phi}({\cal E}_X+p_X) \sum_Y({\cal E}_Y+p_Y)W_Y \ .
\end{align}
%%%%%%%%%%%%%%%
The system of four first-order differential equations [Eqs. \eqref{eq:dK}, \eqref{eq:dH1}, \eqref{eq:dW}, \eqref{eq:dP}] derived above is supplemented by two algebraic relations, which determine the variables $H_{0}$ and $V_{X}$. The algebraic relation for $V_{X}$ follows from the $\theta$-component of the perturbed Euler equation, Eq.~\eqref{eq:perteul}, and takes the form\footnote{The full algebraic derivation is presented in Appendix~\ref{app:eqs}.}
%%%%%%%%%%%%%%%
\begin{align}\label{eq:contV}
    ({\mathcal E}_X+p_X)\omega^2V_X =&\ e^{\Phi}\mathcal P_X-e^{2\Phi-\Lambda}\frac{{\mathcal E}_X+p_X}{r}\Phi'W_X \nonumber\\
    &\ +\frac12({\mathcal E}_X+p_X)e^{2\Phi}H_0 \ ,
\end{align}
%%%%%%%%%%%%%%%
The second algebraic relation, which determines $H_{0}$, follows from an appropriate combination of the $rr$ and $r\theta$ components of the linearized Einstein equations. The resulting expression is given by (see Appendix~\ref{app:eqs} for algebraic details)
%%%%%%%%%%%%%%%
\begin{align}\label{eq:contH}
    &-8\pi re^{2\Lambda-\Phi}\sum_X\mathcal P_X+H_0\left[\left(\frac1r-\Phi'\right)-e^{2\Lambda}\frac{l(l+1)}{2r}\right] \nonumber\\
    &-rK\left[\Phi'\left(\Phi'-\frac1r\right)+e^{2\Lambda-2\Phi}\omega^2-e^{2\Lambda}\frac{(l+2)(l-1)}{2r^2}\right] \nonumber\\
    &+H_1r\left[e^{-2\Phi}\omega^2-\Phi'\frac{l(l+1)}{2r}\right]=0 \ .
\end{align}
%%%%%%%%%%%%%%%
Equations~\eqref{eq:dK}, \eqref{eq:dH1}, \eqref{eq:dW}, and \eqref{eq:dP} define a system of $2N+2$ coupled first-order differential equations for the variables $K$, $H_1$, $W_{X}$, and $\mathcal{P}_{X}$, where $N$ is the number of fluid components. These are supplemented by the algebraic relations \eqref{eq:contV} and \eqref{eq:contH}, which determine $V_{X}$ and $H_{0}$, respectively. Together, these equations form a closed system that can be integrated numerically once appropriate boundary conditions are specified. The resulting formulation constitutes a direct generalization of the single-fluid polar perturbation equations derived in Ref.~\cite{1985ApJ...292...12D} to the case of multiple, gravitationally coupled, non-interacting fluids.
%%%%%%%%%%%%%%%
%%%%%%%%%%%%%%%%%%%%%%%%%%%%%
\subsection{Regularity and boundary conditions}
\label{sec:2d}
%%%%%%%%%%%%%%%%%%%%%%%%%%%%%
%%%%%%%%%%%%%%%
The system of perturbation equations must be supplemented by imposing regularity at the center and boundary conditions at the surfaces of the fluid components. In a multi-fluid configuration, these conditions must be imposed at $r=0$, at each radius $R_{X}$ where a given fluid component vanishes, including the outer stellar surface. Since the system consists of $2N+2$ first-order differential equations, a corresponding set of $2N+2$ boundary conditions is required to obtain a unique solution. We first consider the conditions at the surfaces of the individual fluid components.

At the surface of a given fluid $X$, defined by $p_{X} (R_{X}) = 0$, the appropriate boundary condition is that the Lagrangian pressure perturbation vanishes, i.e. $\Delta_{X}\, p_{X} (R_{X}) = 0$, or equivalently $\mathcal{P}_{X} (R_{X}) = 0$. The remaining perturbation variables, namely $H_{1}$ and $K$, and the fluid variables associated with the other components ($W_{Y},\ \mathcal{P}_{Y}$ with $Y \neq X$), must remain continuous across the surface. For radii $r > R_{X}$, the fluid $X$ is absent, so that $\mathcal{E}_{X} = p_{X} = 0$, and the corresponding variables $W_{X}$ and $\mathcal{P}_{X}$ no longer enter the system of equations. The perturbation problem, therefore, reduces to that of the remaining fluid components in this outer region.

At the stellar center, regularity requires that both the background geometry and the perturbation variables remain finite. To derive the corresponding boundary conditions, we first record the leading behavior of the background quantities near $r=0$. Using Eqs.~\eqref{eq:TOVphi} and \eqref{eq:TOVm}, one finds
%%%%%%%%%%%%%%%
\begin{subequations}
    \begin{align}
        m(r) &\sim \tfrac43\pi r^3 \sum_X\mathcal{E}_X(0)+\mathcal O(r^5)\ , \\
        e^{2\Lambda}(r) &\sim 1+\mathcal O(r^2) \ , \\
        \Phi^{'}(r)&\sim 4\pi r\sum_X[\tfrac{1}{3}\mathcal{E}_X(0)+p_X(0)]+\mathcal O(r^3) \ ,
    \end{align}
\end{subequations}
%%%%%%%%%%%%%%%
so that, in particular, $\Lambda^{'}(r) = \mathcal{O}(r)$ near the center.

With these background scalings, regularity of the perturbation equations implies that the variables entering the first-order system admit regular Taylor expansions about $r = 0$. In particular, one finds
%%%%%%%%%%%%%%%
\begin{subequations}\label{eq:center}
    \begin{align}
        W_X(r) \sim& \ W_X(0)+\mathcal O(r^2)\ , \\
        \mathcal P_X(r) \sim& \ \mathcal P_X(0)+\mathcal O(r^2) \ ,\\
        H_1(r) \sim& \ H_1(0)+\mathcal O(r^2)\ , \\
        K(r) \sim& \ K(0)+\mathcal O(r^2) \ .
    \end{align}
\end{subequations}
%%%%%%%%%%%%%%%
The explicit regularity conditions at the origin are obtained by substituting the expansions of Eq.~\eqref{eq:center} into the perturbation system and requiring that the leading-order terms remain finite. We have checked that the next order terms in this local expansion about the center of the star are negligible for the starting radii chosen in this work. 

We begin with the algebraic relation of Eq.~\eqref{eq:contH}. Its leading contribution near $r=0$ is
%%%%%%%%%%%%%%%
\begin{equation}
    \frac{2-l(l+1)}{2r} H_0(0) - \frac{(l+2)(l-1)}{2r} K(0) + \mathcal O(r^0)=0 \ . \nonumber
\end{equation}
%%%%%%%%%%%%%%%
Since $2-l(l+1) = -(l+2)(l-1)$, regularity for $l\geq2$ requires
%%%%%%%%%%%%%%%
\begin{equation}
    H_{0}(0) = K(0) \ .
    \label{eq:cH0}
\end{equation}
%%%%%%%%%%%%%%%
Next, we evaluate Eq.~\eqref{eq:dW} near the center. Because the expansion \eqref{eq:center} implies $dW/dr = \mathcal{O} (r)$, the singular terms at order $r^{-1}$ must cancel, giving
%%%%%%%%%%%%%%%
\begin{equation}
    -\frac{(l+1)}{r}W_X(0)-\frac{l(l+1)}{r}V_X(0)+\mathcal O(r)=0 \ . \nonumber
\end{equation}
%%%%%%%%%%%%%%%
It follows that regularity at the origin demands
%%%%%%%%%%%%%%%
\begin{equation}
    W_X(0) = -\, l\ V_X(0) \ .
    \label{eq:cV}
\end{equation}
%%%%%%%%%%%%%%%
The remaining central relations are obtained from Eqs.~\eqref{eq:dK} and \eqref{eq:dP}. Using the near-center behavior of the background and perturbation variables and retaining the leading-order terms at $\mathcal{O}(r^{-1})$, Eq.~\eqref{eq:dK} gives
%%%%%%%%%%%%%%%
\begin{align}\label{eq:cH1}
    H_1(0) =&\ \frac{2}{(l+1)}K(0) \nonumber \\
    &\, -\frac{16\pi}{l(l+1)} \sum_X\Big[\mathcal{E}_X(0)+p_X(0)\Big]W_X(0) \ .
\end{align}
%%%%%%%%%%%%%%%
The same relation [Eq.~\eqref{eq:cH1}] is obtained by imposing regularity on Eq.~\eqref{eq:dH1}, providing a useful consistency check of the expansion. Finally, imposing regularity at order $\mathcal{O}(r^{-1})$ in Eq.~\eqref{eq:dP} yields the central value of $\mathcal{P}_{X}$ as
%%%%%%%%%%%%%%%
\begin{align}\label{eq:cPX}
    \mathcal P_X(0) =& \ \Bigg[\Bigg\{\frac{4\pi}{3}\sum_Y\Big(\mathcal{E}_Y(0)+3p_Y(0)\Big) -\frac{\omega^2}{l}e^{-2\Phi(0)}\Bigg\} \nonumber \\
    &  \times W_{X}(0) - \frac{1}{2} K(0)\Bigg]\ e^{\Phi(0)} \big[\mathcal{E}_X(0)+p_X(0)\big] \ .
\end{align}
%%%%%%%%%%%%%%%
An equivalent expression follows by imposing regularity on Eq.~\eqref{eq:contV} at leading order, confirming the consistency of the system.

Together, these relations determine the regular central values of all perturbation variables in terms of the amplitudes $K (0)$ and $W_{X} (0)$, which represent the $N+1$ independent degrees of freedom of the solution at the origin and serve as the starting point for the numerical integration of the system. Imposing the surface conditions $\mathcal{P}_{X}(R_{X}) = 0$ for each fluid component provides $N$ additional constraints. The system is therefore left with one final condition, which determines the allowed eigenfrequencies. This condition is supplied by the behavior of the perturbations in the exterior vacuum region, where the solution must satisfy a purely outgoing-wave boundary condition. The implementation of this exterior matching is described in the next subsection.

%%%%%%%%%%%%%%%%%%%%%%%%%%%%%
%\input{Perturbation_Derivation}
%%%%%%%%%%%%%%%%%%%%%%%%%%%%%
%%%%%%%%%%%%%%%%%%%%%%%%%%%%%
\subsection{Exterior spacetime and matching conditions}
\label{sec:2e}
%%%%%%%%%%%%%%%%%%%%%%%%%%%%%
%%%%%%%%%%%%%%%
Outside the outermost fluid surface, $r > R_{\rm{out}}$, the matter fields vanish and the background spacetime reduces to the Schwarzschild solution with mass $M = m(R_{\rm{out}})$. In this exterior region, the perturbation problem reduces to that of even-parity vacuum metric perturbations of a Schwarzschild spacetime. The interior solution obtained by integrating the coupled two-fluid perturbation equations must therefore be matched at $r = R_{\rm{out}}$ to a corresponding exterior vacuum solution. Once the perturbation variables are specified at the matching surface, the subsequent determination of the mode spectrum follows the standard procedure for vacuum perturbations of Schwarzschild spacetime. For completeness, we outline this procedure below, closely following the established treatments in the literature~\cite{PhysRevD.65.024010, PhysRevD.102.063025}. The remaining condition required to determine the eigenfrequencies is then provided by the asymptotic behavior of the perturbations at large distances, where physically admissible solutions must satisfy a purely outgoing-wave condition.

In the exterior vacuum region, the even-parity metric perturbations can be expressed in terms of a single gauge-invariant master function, the Zerilli function $Z(r)$, which satisfies a second-order wave equation on the Schwarzschild background. Introducing the tortoise coordinate $r_{*} = r + 2M \ln(r/2M - 1)$, the perturbation equations reduce to a Schrödinger-like equation of the form~\cite{PhysRevLett.24.737}
%%%%%%%%%%%%%%%
\begin{align}
    \frac{d^{2}Z}{dr_{*}^{2}} + \left(\omega^{2} - V_{Z}\right) Z = 0\ , 
\end{align}
%%%%%%%%%%%%%%%
where $V_{Z} (r)$ is the Zerilli potential, given by 
%%%%%%%%%%%%%%%
\begin{align}
    & V_{Z} = \left(1 - \frac{2M}{r}\right) \times \nonumber \\
    & \left(\frac{2 \lambda^{2}(\lambda+1)r^{3} + 6\lambda^{2}Mr^{2} + 18\lambda M^{2} r + 18 M^{3}}{r^{3} \left(\lambda r + 3M\right)^{2}}\right) \ ,
\end{align}
%%%%%%%%%%%%%%%
with $\lambda \equiv [(l-1)(l+2)]/2$. The exterior even-parity solution is therefore fully determined once $Z$ and its radial derivative are specified at the matching surface.

At the matching surface, i.e., $r = R_{\rm{out}}$, the interior solution provides the metric perturbation functions $K (R_{\rm{out}})$ and $H_{1} (R_{\rm{out}})$. To connect this interior solution to the exterior Zerilli formulation, these variables must be expressed in terms of the Zerilli function $Z$ and its derivative. In the exterior region, the metric perturbations can be written in terms of $Z$ through the standard relations~\cite{1975RSPSA.344..441C, 1983ApJS...53...73L}, which allow one to determine $Z$ and $dZ/dr_{*}$ at the stellar surface ($R_{\rm{out}}$) from the interior solution. These relations take the form
%%%%%%%%%%%%%%%
\begin{subequations}\label{eq:zerilli_transform}
\begin{align}
    r^{l} K &= \frac{\lambda(\lambda+1)r^{2}+3\lambda r M + 6M^{2}}{r^{2}(\lambda r+3M)}Z + \frac{dZ}{dr_{*}} \\
    r^{l+1} H_{1} &= \frac{\lambda r^{2} - 3\lambda rM - 3M^{2}}{(r-2M)(\lambda r+3M)}Z + \frac{r^{2}}{r-2M} \frac{dZ}{dr_{*}}\,,
\end{align}
\end{subequations}
%%%%%%%%%%%%%%%
which are to be evaluated at the outer stellar surface, $r = R_{\rm{out}}$. In this way, the metric perturbation at the stellar surface, $K (R_{\rm{out}})$ and $H_{1} (R_{\rm{out}})$, determine the corresponding exterior values $Z (R_{\rm{out}})$ and $dZ/dr_{*} (R_{\rm{out}})$, thereby providing the necessary initial data for the exterior problem. 

Since we are considering even-parity perturbations, the exterior spacetime is naturally described by the Zerilli function $Z$. For the quasi-normal-mode calculation, however, we transform the Zerilli equation to the Regge–Wheeler form using the Chandrasekhar transformation~\cite{1975RSPSA.344..441C}. We adopt this representation because the Regge–Wheeler potential is simpler, which makes the asymptotic analysis and the continued-fraction implementation more transparent, while leaving the resulting eigenfrequencies unchanged (since isospectrality is preserved here). In terms of the Regge–Wheeler function $Q(r)$, the exterior perturbations satisfy~\cite{PhysRev.108.1063}
%%%%%%%%%%%%%%%
\begin{align}\label{eq:Regge-Wheeler}
    \frac{d^{2}Q}{dr_{*}^{2}} + \left(\omega^{2} - V_{\rm{RW}}\right)Q = 0 \ ,
\end{align}
%%%%%%%%%%%%%%%
where the Regge–Wheeler potential $V_{\rm{RW}}$ is given by
%%%%%%%%%%%%%%%
\begin{align}
    V_{\rm{RW}} = \left(1 - \frac{2M}{r}\right) \left(\frac{l(l+1)}{r^{2}} - \frac{6M}{r^{3}}\right)\ .
\end{align}
%%%%%%%%%%%%%%%
The Zerilli and Regge–Wheeler functions are related through the Chandrasekhar transformation~\cite{1975RSPSA.344..441C, PhysRevD.48.3467}, which takes the form
%%%%%%%%%%%%%%%
\begin{subequations}\label{eq:Chandrasekhar_transform}
\begin{align}
    &\left(\kappa + 2i\omega\beta\right)Z = \left(\kappa + \beta^{2} \frac{r-2M}{r^{2}(\lambda r + 3M)}\right)Q + 2\beta \frac{dQ}{dr_{*}} \\
    &\left(\kappa - 2i\omega\beta\right)Q = \left(\kappa + \beta^{2} \frac{r-2M}{r^{2}(\lambda r + 3M)}\right)Z - 2\beta \frac{dZ}{dr_{*}}   
\end{align}
\end{subequations}
%%%%%%%%%%%%%%%
with $\beta = 6M$ and $\kappa = 4\lambda(\lambda+1)$. This representation is particularly convenient for imposing the outgoing-wave boundary condition.

To determine the allowed eigenfrequencies, one must impose the appropriate asymptotic behavior on the exterior solution. In the limit $r \rightarrow \infty$, the Regge–Wheeler potential vanishes, and the exterior solution approaches a superposition of ingoing and outgoing waves. With the time dependence $e^{i\omega t}$ adopted throughout this work, the physically admissible quasi-normal mode solution is purely outgoing and behaves as $e^{-i\omega r_{*}}$ at spatial infinity, corresponding to $e^{i\omega(t-r_{*})}$. To implement this condition, we expand the Regge–Wheeler function in a form that explicitly factors out the required asymptotic behavior.

Following the standard continued-fraction construction, the exterior solution is expanded about a finite radius $r_{a}$ in the exterior vacuum region, chosen so that the series converges~\cite{10.1046/j.1365-8711.1999.02983.x}. In practice, we take $r_{a} = R_{\rm out}$ when $R_{\rm out} \geq 4M$, since the stellar surface itself then provides a suitable expansion point for the continued-fraction series. For more compact configurations with $R_{\rm out}<4M$, we instead set $r_{a} = 8M$, so that the expansion point is moved farther out in the vacuum exterior and the series remains convergent up to spatial infinity. When $r_a \neq R_{\rm out}$, the exterior solution is first propagated from the stellar surface to the expansion radius. The interior solution determines $K$ and $H_1$ at $R_{\rm out}$, from which $Z$ and $dZ/dr_*$ are obtained using Eqs.~\eqref{eq:zerilli_transform}. The Chandrasekhar transformation [Eq.~\eqref{eq:Chandrasekhar_transform}] is then used to construct $Q$ and $dQ/dr_*$ at $R_{\rm out}$, and the Regge--Wheeler equation is integrated outward through the vacuum region up to $r = r_a$. The resulting $Q(r_a)$ and $(dQ/dr)|_{r_a}$ are used to initialize the continued-fraction expansion. We verified that shifting $r_{a}$ within the exterior region does not change the extracted real mode frequencies to the numerical precision of the present calculation. 

The exterior solution is then written as
%%%%%%%%%%%%%%%
\begin{align}
    Q(r) = e^{-i\omega r_{*}} \sum_{n=0}^{\infty} a_{n} \left(1 - \frac{r_{a}}{r}\right)^{n}\ .
\end{align}
%%%%%%%%%%%%%%%
Substituting the series representation above into the Regge–Wheeler equation and collecting terms order by order in $(1 - r_{a}/r)$ yields a recurrence relation for the expansion coefficients $a_{n}$. For $n\geq 2$, the coefficients satisfy the four-term relation (one could also obtain a 3-term recursion relation directly with a different form of the continued fraction ansatz~\cite{Karikos:2026arz})
%%%%%%%%%%%%%%%
\begin{align}
    \alpha_{n}a_{n+1} + \beta_{n}a_{n} + \gamma_{n}a_{n-1} + \delta_{n}a_{n-2} = 0\ ,
\end{align}
%%%%%%%%%%%%%%%
where the recurrence coefficients are given by
%%%%%%%%%%%%%%%
\begin{subequations}
\begin{align}
    \alpha_{n} &= \left(1 - \frac{2M}{r_{a}}\right) n (n+1)\ , \\
    \beta_{n} &= -2\left[i\omega r_{a} + \left(1 - \frac{3M}{r_{a}}\right)n\right]n\ , \\
    \gamma_{n} &= \left(1 - \frac{6M}{r_{a}}\right)n(n-1) + \frac{6M}{r_{a}} - l(l+1)\ , \\
    \delta_{n} &= \frac{2M}{r_{a}} (n-3) (n+1)\ .
\end{align}
\end{subequations}
%%%%%%%%%%%%%%%
To initialize the recurrence, the first coefficients must be specified separately. The structure of the series expansion implies $a_{-1} = 0$, while the coefficients $a_{0}$ and $a_{1}$ are determined by the values of $Q$ and its radial derivative with respect to $r$ at $r = r_{a}$. These coefficients are given by
%%%%%%%%%%%%%%%
\begin{subequations}
\begin{align}
    a_{0} &= \frac{Q(r_{a})}{e^{-i\omega r_{*}}|_{r = r_{a}}} \ ,\\
    a_{1} &= \frac{r_{a}}{e^{-i\omega r_{*}}|_{r = r_{a}}} \left[\frac{dQ}{dr}\Bigg|_{r_{a}} + \frac{i\omega r_{a}}{r_{a}-2M}Q(r_{a})\right] \ .
\end{align}
\end{subequations}
%%%%%%%%%%%%%%%
For the continued-fraction construction, it is convenient to reduce the four-term recurrence relation to a three-term form. This can be achieved through a standard Gaussian-elimination procedure applied to the recurrence relation. After Gaussian elimination, the resulting three-term recurrence involves coefficients $\hat{\alpha}_{n}$, $\hat{\beta}_{n}$ and $\hat{\gamma}_{n}$, which we define recursively by:
%%%%%%%%%%%%%%%
\begin{equation}
    \hat{\alpha}_{1} = \alpha_1\, , \quad \quad \hat{\beta}_{1} = \beta_1\, , \quad \quad \hat{\gamma}_{1} = \gamma_1\, ,
\end{equation}
%%%%%%%%%%%%%%%
 and, for $n \geq 2$,
%%%%%%%%%%%%%%%
\begin{subequations}
\begin{align}
    \hat{\alpha}_{n} &= \alpha_{n}\, ,  \\
    \hat{\beta}_{n} &= \beta_{n} - \frac{\hat{\alpha}_{n-1}\delta_{n}}{\hat{\gamma}_{n-1}}\, ,  \\
    \hat{\gamma}_{n} &= \gamma_{n} - \frac{\hat{\beta}_{n-1}\delta_{n}}{\hat{\gamma}_{n-1}}\, .
\end{align}
\end{subequations}
%%%%%%%%%%%%%%%
The four-term recurrence relation can be recast into the three-term form~\cite{PhysRevD.41.2986}
%%%%%%%%%%%%%%%
\begin{align}
    \hat{\alpha}_{n}a_{n+1} + \hat{\beta}_{n}a_{n} + \hat{\gamma}_{n}a_{n-1} = 0\ .
\end{align}
%%%%%%%%%%%%%%%
After reducing the recurrence relation to the three-term form, the quasi-normal mode condition can be imposed through the associated continued fraction. In particular, requiring the sequence of coefficients $a_{n}$ to define the minimal solution of the recurrence leads to a continued-fraction equation for the frequency $\omega$ (Leaver’s method)~\cite{PhysRevD.48.3467}, given by
%%%%%%%%%%%%%%%
\begin{align}
{\cal F}(\omega) \equiv \frac{a_1}{a_0}
+ \cfrac{\hat{\gamma}_1}{\hat{\beta}_1
- \cfrac{\hat{\alpha}_1\hat{\gamma}_2}{\hat{\beta}_2
- \cfrac{\hat{\alpha}_2\hat{\gamma}_3}{\hat{\beta}_3-\cdots}}} = 0 \ .
\end{align}
%%%%%%%%%%%%%%%
Here, the coefficients $\hat{\alpha}_{n}$, $\hat{\beta}_{n}$, and $\hat{\gamma}_{n}$ of the three-term recurrence relation are determined entirely by the exterior Regge--Wheeler problem, whereas the dependence on the stellar interior enters through the ratio $a_{1}/a_{0}$, fixed by the values of $Q$ and $dQ/dr$ at the expansion radius $r_a$. The allowed eigenfrequencies are determined by the roots of ${\cal F}(\omega) = 0$.

%%%%%%%%%%%%%%%
%%%%%%%%%%%%%%%%%%%%%%%%%%%%%
\section{Numerical implementation}
\label{sec:3}
%%%%%%%%%%%%%%%%%%%%%%%%%%%%%
%%%%%%%%%%%%%%%
In this section, we describe the numerical procedure used to solve the perturbation equations derived in Sec.~\ref{sec:2}. Although the formalism was developed for a general multi-fluid system, the numerical construction presented here is restricted to the case of two gravitationally coupled fluids. In what follows, we label these components as $I$ and $O$, denoting the inner and outer fluids, respectively. For a given choice of equations of state, equilibrium configurations are first constructed by solving the background equations outlined in Sec.~\ref{sec:2a}. Each fluid has its own surface, determined by the vanishing of its pressure, and the corresponding radii are denoted by $R_{I}$ and $R_{O}$. In general, the two radii need not coincide, although they may do so in special cases. We choose the labels such that $R_{I} \leq R_{O}$, so that the surface of the outer fluid component defines the outer stellar boundary, $R_{\rm{out}} = R_{O}$. The numerical matching construction described below therefore applies to the configurations with $R_{I} < R_{O}$ considered here. The resulting background profiles, including the metric functions and thermodynamic variables of both fluids, are then used as input for the numerical integration of the perturbation equations.

The perturbation equations define a linear boundary-value problem for each trial frequency $\omega$. For two fluids, the first-order system contains six dynamical variables, and hence six boundary conditions are required. Regularity at the center provides three relations among the central values, leaving three independent central amplitudes. The two fluid-surface conditions $\mathcal{P}_{I} (R_{I}) = 0$ and $\mathcal{P}_{O}(R_{O}) = 0$ provide two additional conditions, and the final condition is the purely outgoing-wave condition imposed on the exterior solution. The regularity conditions at the center do not select a unique solution, because the central expansion still contains free amplitudes. Similarly, imposing the surface boundary conditions does not by itself determine a unique global solution, since additional free amplitudes remain in the solutions constructed from the outer surface. We therefore solve the problem by constructing independent basis solutions that satisfy the admissible local conditions at the center and at the relevant fluid surfaces, and then determine the physical solution by matching these basis solutions at an intermediate radius.

%%%%%%%%%%%%%%%%%%%%%%%%%%%%%%%%%%%%%%%%%%%%%%%%%%%%%%%%%%
\begin{figure*}[tbp]
\centering
\begin{tikzpicture}[
    font=\normalsize,
    >=Latex,
    box/.style={
        rectangle,
        rounded corners=5pt,
        draw=#1!0!black,
        fill=white,
        line width=1.5pt,
        align=center,
        inner sep=5pt,
        minimum height=1.25cm
    },
    box/.default=blue,
    arrow/.style={
        -{Latex[length=2.0mm]},
        line width=1.00pt
    },
    dashedarrow/.style={
        -{Latex[length=2.0mm]},
        dashed,
        line width=1.00pt
    }
]

\tikzset{flowarrow/.style={-{Stealth[length=3.4mm,width=2.4mm]},
    line width=1.5pt, draw=blue}}

\tikzset{
    arrowlabel/.style={
        font=\normalsize,
        fill=white,
        inner sep=1pt
    }
}

% ------------------------------------------------------------
% Top row: background and trial frequency
% ------------------------------------------------------------

\node[box=gray, text width=12.0cm, minimum height=2.05cm] (bg) at (0.0, 0) {};
\fill[gray!25, rounded corners=4pt]
    ([xshift=1.5pt,yshift=-1.5pt]bg.north west)
    rectangle
    ([xshift=-1.5pt,yshift=-0.60cm]bg.north east);
\draw[dashed, line width=0.50pt, gray!60!black]
    ([xshift=1.5pt,yshift=-0.60cm]bg.north west)
    --
    ([xshift=-1.5pt,yshift=-0.60cm]bg.north east);

\node[align=center] at ([yshift=-0.30cm]bg.north) {
\tikz[baseline=(num.base)]{
  \node[circle, fill=black!100, text=white, inner sep=0pt, minimum size=3.75mm] (num) {\bfseries 1};
}
\hspace{0.01em}\textbf{Two-fluid background}
};

\node[align=center, text width=11.6cm] at ([yshift=-1.35cm]bg.north) {
Choose the equations of state and specify central densities for two fluids.\\
Solve the equilibrium structure to obtain profiles for $\Phi (r)$, $\Lambda (r)$, ${\cal E}_{I,O} (r)$, $p_{I,O} (r)$, and the fluid surfaces $R_{I}$ and $R_{O}$, with $R_I\leq R_O$.
};

\node[box=blue, text width=6.85cm, minimum height=1.55cm] (trial) at (0,-2.5) {};
\fill[gray!25, rounded corners=4pt]
    ([xshift=1.5pt,yshift=-1.5pt]trial.north west)
    rectangle
    ([xshift=-1.5pt,yshift=-0.60cm]trial.north east);
\draw[dashed, line width=0.50pt, gray!60!black]
    ([xshift=1.5pt,yshift=-0.60cm]trial.north west)
    --
    ([xshift=-1.5pt,yshift=-0.60cm]trial.north east);

\node[align=center] at ([yshift=-0.30cm]trial.north) {
\tikz[baseline=(num.base)]{
  \node[circle, fill=black!100, text=white, inner sep=0.0pt, minimum size=3.75mm] (num) {\bfseries 2};
}
\hspace{0.01em}\textbf{Trial frequency and matching radius $\mathbf{r_{\rm{\bf m}}}$}
};

\node[align=center, text width=6.50cm] at ([yshift=-1.10cm]trial.north) {
Select a trial $\omega$ and set $r_{\rm{m}} = (R_{I} + R_{O})/2$.
\textsuperscript{*}};

% ------------------------------------------------------------
% Informational note connected to box 2
% ------------------------------------------------------------
%\node[
%    rectangle,
%    rounded corners=6pt,
%    draw=gray!70!black,
%    dashed,
%    line width=1.50pt,
%    fill=gray!5,
%    align=left,
%    text width=3.70cm,
%   inner sep=6pt,
%    font=\footnotesize
%] (rmnote) at (6.75,-2.50) {
%\textit{$r_{\rm m}$ is an auxiliary interior matching point chosen in the outer-fluid region $R_I < r_{\rm m} < R_O$; the extracted frequencies are independent of its precise location within this region.}
%};

%\draw[dashedarrow] (trial.east) -- (rmnote.west);

\draw[flowarrow] (bg.south) -- ++(0, -0.45) -| (trial.north);

% ------------------------------------------------------------
% Integration branches
% ------------------------------------------------------------

\node[box=gray, text width=8.2cm, minimum height=3.50cm] (center) at (-4.5,-6.0) {};
\fill[gray!25, rounded corners=4pt]
    ([xshift=1.5pt,yshift=-1.5pt]center.north west)
    rectangle
    ([xshift=-1.5pt,yshift=-0.60cm]center.north east);
\draw[dashed, line width=0.50pt, gray!60!black]
    ([xshift=1.5pt,yshift=-0.60cm]center.north west)
    --
    ([xshift=-1.5pt,yshift=-0.60cm]center.north east);

\node[align=left, text width=8.2cm] at ([yshift=-0.30cm]center.north) {
\tikz[baseline=(num.base)]{
  \node[circle, fill=black!100, text=white, inner sep=0pt, minimum size=4.0mm] (num) {\bfseries 3a};
}
\hspace{0.01em}\textbf{Center $\rightarrow$ outward integration}
};

\node[align=left, text width=8.2cm] at ([yshift=-2.01cm]center.north) {
\begin{itemize}[leftmargin=*, labelsep=0.4em, labelwidth=0.8em,
    itemindent=0em, topsep=2pt, parsep=0pt, partopsep=0pt, itemsep=1pt]
    \item Use regularity at $r = 0$.
    \item Choose two central data sets: $K(0)=\pm[{\cal E}(0)+p(0)]$, with $W_{I}(0)=1$ in each case.
    \item Adjust shooting parameter $W_O(0)$ until ${\cal P}_I(R_I)=0$.
    \item This yields two regular basis solutions, integrated outward to $r_{\rm m}$.
\end{itemize}
};

\node[box=gray, text width=8.2cm, minimum height=3.50cm] (outer) at (4.5,-6.0) {};
\fill[gray!25, rounded corners=4pt]
    ([xshift=1.5pt,yshift=-1.5pt]outer.north west)
    rectangle
    ([xshift=-1.5pt,yshift=-0.60cm]outer.north east);
\draw[dashed, line width=0.50pt, gray!60!black]
    ([xshift=1.5pt,yshift=-0.60cm]outer.north west)
    --
    ([xshift=-1.5pt,yshift=-0.60cm]outer.north east);

\node[align=left, text width=8.2cm] at ([yshift=-0.30cm]outer.north) {
\tikz[baseline=(num.base)]{
  \node[circle, fill=black!100, text=white, inner sep=0pt, minimum size=4.0mm] (num) {\bfseries 3b};
}
\hspace{0.01em}\textbf{Outer surface $\rightarrow$ inward integration}
};

\node[align=left, text width=8.2cm] at ([yshift=-2.01cm]outer.north) {
\begin{itemize}[leftmargin=*, labelsep=0.4em, labelwidth=0.8em,
    itemindent=0em, topsep=2pt, parsep=0pt, partopsep=0pt, itemsep=1pt]
    \item Use reduced perturbation vector $\mathbf{\Psi}(r)=(H_1,K,W_O,{\cal P}_O)^T$.
    \item Impose ${\cal P}_O(R_O)=0$ and choose three independent basis $(H_1,K,W_O)=(1,0,0)$, $(0,1,0)$, and $(0,0,1)$.
    \item Integrate the reduced perturbation system inward to $r_{\rm m}$; this yields three inward-integrated basis solutions.
\end{itemize}
};

\draw[flowarrow] (trial.south) -- ++(0,-0.45) -| (center.north);
\draw[flowarrow] (trial.south) -- ++(0,-0.45) -| (outer.north);

% ------------------------------------------------------------
% Matching
% ------------------------------------------------------------

\node[box=gray, text width=11.0cm, minimum height=2.05cm] (match) at (0,-10.40) {};
\fill[gray!25, rounded corners=4pt]
    ([xshift=1.5pt,yshift=-1.5pt]match.north west)
    rectangle
    ([xshift=-1.5pt,yshift=-0.60cm]match.north east);
\draw[dashed, line width=0.50pt, gray!60!black]
    ([xshift=1.5pt,yshift=-0.60cm]match.north west)
    --
    ([xshift=-1.5pt,yshift=-0.60cm]match.north east);

\node[align=center, text width=10.2cm] at ([yshift=-0.30cm]match.north) {
\tikz[baseline=(num.base)]{
  \node[circle, fill=black!100, text=white, inner sep=0pt, minimum size=3.75mm] (num) {\bfseries 4};
}
\hspace{0.01em}\textbf{Match the basis solutions at $\mathbf{r}$ $\mathbf{=}$ $\mathbf{r}_{\rm \bf m}$}
};

\node[align=center, text width=10.2cm] at ([yshift=-1.36cm]match.north) {
Require continuity: $\mathbf{\Psi}_{\rm in}(r_{\rm m})=\mathbf{\Psi}_{\rm out}(r_{\rm m})$\\
Four equations for five coefficients $a_1,\ldots,a_5$\\
Fix one overall normalization and solve the resulting $4\times4$ system
};

\coordinate (caux) at ($(center.south)+(0,-0.45)$);
\coordinate (oaux) at ($(outer.south)+(0,-0.45)$);

\coordinate (caux) at ($(center.south)+(0,-0.45)$);
\coordinate (oaux) at ($(outer.south)+(0,-0.45)$);

\draw[flowarrow] (center.south) -- node[arrowlabel,pos=1.70,left=-40.0pt] {$\mathbf{\Psi}_{\rm in} = a_1 \mathbf{\Psi}_{\rm in}^{(1)} + a_2 \mathbf{\Psi}_{\rm in}^{(2)}$} (caux);
\draw[flowarrow] (caux) -| (match.north);

\draw[flowarrow] (outer.south) -- node[arrowlabel,pos=1.70,right=-40.0pt] {$\mathbf{\Psi}_{\rm out} = a_3 \mathbf{\Psi}_{\rm out}^{(1)} + a_4 \mathbf{\Psi}_{\rm out}^{(2)} + a_5 \mathbf{\Psi}_{\rm out}^{(3)}$} (oaux);
\draw[flowarrow] (oaux) -| (match.north);

% ------------------------------------------------------------
% Middle output row
% ------------------------------------------------------------
\node[box=gray, fill=gray!25, align=left, text width=7.0cm, minimum height=1.50cm] (global) at (-5.0,-13.25) {
\tikz[baseline=(num.base)]{
  \node[circle, fill=black!100, text=white, inner sep=0pt, minimum size=3.75mm] (num) {\bfseries 5};
}
\hspace{0.01em} \textbf{Construct a single global interior solution that is regular at the center, continuous at $\mathbf{r_{\rm \bf{m}}}$, and satisfies the fluid-surface conditions.}
};

\node[box=gray, text width=7.0cm, minimum height=1.50cm] (ext) at (5.0,-13.25) {};
\fill[gray!25, rounded corners=4pt]
    ([xshift=1.5pt,yshift=-1.5pt]ext.north west)
    rectangle
    ([xshift=-1.5pt,yshift=-0.60cm]ext.north east);
\draw[dashed, line width=0.50pt, gray!60!black]
    ([xshift=1.5pt,yshift=-0.60cm]ext.north west)
    --
    ([xshift=-1.5pt,yshift=-0.60cm]ext.north east);

\node[align=center, text width=7.20cm] at ([yshift=-0.30cm]ext.north) {
\tikz[baseline=(num.base)]{
  \node[circle, fill=black!100, text=white, inner sep=0pt, minimum size=3.75mm] (num) {\bfseries 6};
}
\hspace{0.01em}\textbf{Initialize the exterior problem at $\mathbf r$ $\mathbf =$ $\mathbf{R_{O}}$}
};

\node[align=center, text width=7.20cm] at ([yshift=-1.08cm]ext.north) {
Extract $H_1(R_O)$ and $K(R_O)$; and map to the Zerilli--Regge--Wheeler representation
};

\draw[flowarrow] (match.south) -- ++(0,-0.45) -| (global.north);
\draw[flowarrow] (global.east) -- (ext.west);

% ------------------------------------------------------------
% Final row
% ------------------------------------------------------------
\node[box=gray, text width=7.2cm, minimum height=2.10cm] (resid) at (-5.0,-16.00) {};
\fill[gray!25, rounded corners=4pt]
    ([xshift=1.5pt,yshift=-1.5pt]resid.north west)
    rectangle
    ([xshift=-1.5pt,yshift=-0.60cm]resid.north east);
\draw[dashed, line width=0.35pt, gray!70!black]
    ([xshift=1.5pt,yshift=-0.60cm]resid.north west)
    --
    ([xshift=-1.5pt,yshift=-0.60cm]resid.north east);

\node[align=center, text width=6.55cm] at ([yshift=-0.30cm]resid.north) {
\tikz[baseline=(num.base)]{
  \node[circle, fill=black!100, text=white, inner sep=0pt, minimum size=3.75mm] (num) {\bfseries 8};
}
\hspace{0.01em}\textbf{Mode identification}
};

\node[align=center, text width=7.20cm] at ([yshift=-1.38cm]resid.north) {
Locate the deep minima of $|{\cal F}(\omega)|$ to obtain mode frequencies, and extract the corresponding eigenfunctions.
};

\node[box=gray, text width=7.20cm, minimum height=2.10cm] (modes) at (5.0,-16.00) {};
\fill[gray!25, rounded corners=4pt]
    ([xshift=1.5pt,yshift=-1.5pt]modes.north west)
    rectangle
    ([xshift=-1.5pt,yshift=-0.60cm]modes.north east);
\draw[dashed, line width=0.35pt, gray!70!black]
    ([xshift=1.5pt,yshift=-0.60cm]modes.north west)
    --
    ([xshift=-1.5pt,yshift=-0.60cm]modes.north east);

\node[align=center, text width=7.20cm] at ([yshift=-0.30cm]modes.north) {
\tikz[baseline=(num.base)]{
  \node[circle, fill=black!100, text=white, inner sep=0pt, minimum size=3.75mm] (num) {\bfseries 7};
}
\hspace{0.01em}\textbf{Continued-fraction residual}
};

\node[align=center, text width=7.20cm] at ([yshift=-1.38cm]modes.north) {
For the chosen $\omega$, evaluate ${\cal F}(\omega)$ from exterior solution
and repeat the calculation over a range of frequencies to construct $|{\cal F}(\omega)|$.
};

\draw[flowarrow] (ext.south) -- ++(0,-0.45) -| (modes.north);
\draw[flowarrow] (modes.west) -- (resid.east);

\end{tikzpicture}
\caption{Flowchart of the numerical procedure used to construct the non-radial mode spectrum of a gravitationally coupled two-fluid star. The notation $I$ and $O$ denote the inner and outer fluid components, respectively. \textsuperscript{*}The matching point is chosen in the outer-fluid region, $R_I < r_{\rm m} <R_O$, where the reduced matching vector $\mathbf{\Psi}=(H_1,K,W_O,{\cal P}_O)^T$ is appropriate. The extracted frequencies are independent of the precise location of $r_{\rm m}$ within this region.}
\label{fig:numerical_flowchart}
\end{figure*}
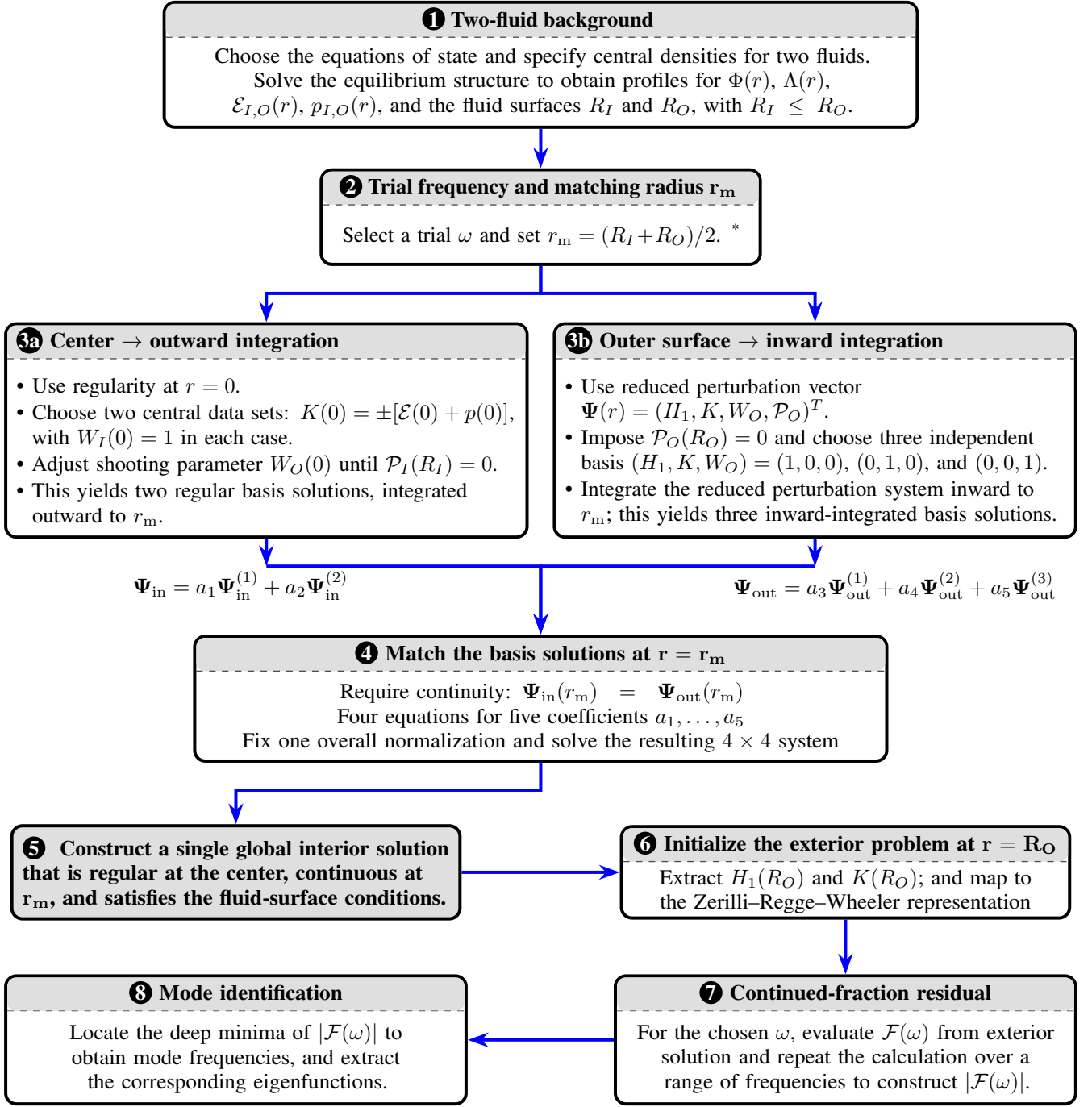
%%%%%%%%%%%%%%%%%%%%%%%%%%%%%%%%%%%%%%%%%%%%%%%%%%%%%%%%%%

For a given trial frequency $\omega$, the perturbation equations [Eqs.~\eqref{eq:dK}, \eqref{eq:dH1}, \eqref{eq:dW}, \eqref{eq:dP}] are first integrated outward from the stellar center. The regularity conditions derived in Sec.~\ref{sec:2d} determine the central values of the perturbation variables in terms of the free amplitudes $K(0)$, $W_I(0)$, and $W_O(0)$ [Eqs.~\eqref{eq:cH0}$-$\eqref{eq:cPX}], where recall that $I$ and $O$ denote the inner and outer fluids, respectively. To construct a convenient pair of independent regular interior basis solutions, we choose the central data $(K(0),\ W_I(0))=(\pm[{\cal E}(0)+p(0)],\ 1)$\footnote{The choice $W_{I}(0) = 1$ is a normalization used to construct a convenient pair of independent regular solutions satisfying $\mathcal{P}_{I}(R_{I}) = 0$. Since the perturbation equations are linear, any solution with $W_{I}(0) \neq 0$ can be rescaled to this form. An exactly inner-displacement-free solution with $W_{I}(0) = 0$, if present, would require constructing the full three-dimensional regular center basis and imposing $\mathcal{P}_{I}(R_{I}) = 0$ by linear algebra.}, where ${\cal E}(0)\equiv{\cal E}_I(0)+{\cal E}_O(0)$ and $p(0)\equiv p_I(0)+p_O(0)$ denote the total central energy density and pressure of the two-fluid background configuration. For each of these two choices, the remaining central amplitude $W_O(0)$ is treated as a shooting parameter and adjusted until the inner-fluid surface condition ${\cal P}_I(R_I)=0$ is satisfied. This procedure yields two linearly independent interior solutions that are regular at the center and satisfy the boundary condition at the inner-fluid surface, and these solutions are then integrated outward from the center to the intermediate matching radius.

The matching radius $r_{\rm m}$ is an auxiliary point at which the independently constructed outward- and inward-integrated interior basis solutions are matched. In the implementation described here, $r_{\rm m}$ is chosen in the outer-fluid region, $R_{I} < r_{\rm m} < R_{O}$, where the inner fluid is absent and the reduced perturbation vector $\mathbf{\Psi}(r) = (H_{1}(r), K(r), W_{O}(r), {\cal P}_{O}(r))^{T}$ is appropriate\footnote{If the matching point were chosen in the two-fluid region, $0<r_{\rm m}<R_I$, the matching vector would have to be enlarged to include the inner-fluid variables \(W_I\) and \({\cal P}_I\). The reduced four-dimensional vector used in the present implementation is therefore tied to the choice $R_I<r_{\rm m}<R_O$.}. The computed mode frequencies are independent of the precise location of $r_{\rm m}$, provided it is varied within this regular matching region. For the calculations presented here, we choose $r_{\rm m} = (R_{I} + R_{O})/2$. In this region, the variables associated with fluid $I$ no longer enter the perturbation equations, and the system reduces to the metric variables together with the perturbation variables of the outer fluid $O$. For the same trial frequency $\omega$, we then construct a second set of independent solutions by integrating the reduced perturbation system inward from the outer stellar surface $r = R_{O}$ to the matching radius $r = r_{\rm m}$. At $r = R_{O}$, we impose the outer-fluid surface condition ${\cal P}_{O}(R_{O}) = 0$, while the remaining variables $H_{1}(R_{O})$, $K(R_{O})$, and $W_{O}(R_{O})$ are chosen freely to generate a basis of three independent inward-integrated solutions. In practice, we use the three basis choices $(H_{1}, K, W_{O}) = (1, 0, 0),\ (0, 1, 0)$, and $(0, 0, 1)$ at $r = R_{O}$, with ${\cal P}_{O}(R_{O}) = 0$ in each case. These three solutions are integrated inward to $r_{\rm m}$, where they are matched to the two outward-integrated interior basis solutions constructed from the center.

To make this matching construction explicit, we introduce the relevant perturbation vector to be evaluated at $r = r_{\rm m}$,
%%%%%%%%%%%%%%%
\begin{align}
    \mathbf{\Psi}(r) = (H_{1}(r), K(r), W_{O}(r), {\cal P}_{O}(r))^{T}\ ,
\end{align}
%%%%%%%%%%%%%%%
which contains the independent metric and outer-fluid variables of the reduced perturbation system in the region $R_{I} < r < R_{O}$. For a fixed trial frequency $\omega$, the general outward-integrated solution in this region can be written as a linear combination of the two regular interior basis solutions,
%%%%%%%%%%%%%%%
\begin{align}
    \mathbf{\Psi}_{\rm{in}}(r) = a_{1} \mathbf{\Psi}_{\rm{in}}^{(1)}(r) + a_{2} \mathbf{\Psi}_{\rm{in}}^{(2)}(r)\ .
\end{align}
%%%%%%%%%%%%%%%
Here $a_{1}$ and $a_{2}$ are matching coefficients to be determined, while the superscripts (1) and (2) label the two center-regular basis solutions constructed from the central choices $K(0) = +[{\cal E}(0)+ p(0)]$ and $K(0) = -[{\cal E}(0)+ p(0)]$, respectively, with $W_{I}(0) = 1$ in both cases and $W_{O}(0)$ fixed by ${\cal P}_{I}(R_{I}) = 0$. For the inward-integrated solutions, the general solution in the same outer-fluid region is written as a linear combination of the three basis solutions constructed from the outer surface,
%%%%%%%%%%%%%%%
\begin{align}
    \mathbf{\Psi}_{\rm{out}}(r) = \ a_{3} \mathbf{\Psi}_{\rm{out}}^{(1)}(r) + a_{4} \mathbf{\Psi}_{\rm{out}}^{(2)}(r) + a_{5} \mathbf{\Psi}_{\rm{out}}^{(3)}(r)\ .
\end{align}
%%%%%%%%%%%%%%%
Here $a_{3}$, $a_{4}$, and $a_{5}$ are additional matching coefficients, and the three superscripts label the basis solutions generated by the choices $(H_{1}, K, W_{O}) = (1, 0, 0),\ (0, 1, 0)$, and $(0, 0, 1)$ at $r = R_{O}$, with ${\cal P}_{O}(R_{O}) = 0$. Continuity of the physical solution at the matching radius requires the outward- and inward-integrated linear combinations to agree, so that
%%%%%%%%%%%%%%%
\begin{align}
    a_{1} \mathbf{\Psi}_{\rm{in}}^{(1)} (r_{\rm m}) + a_{2} \mathbf{\Psi}_{\rm{in}}^{(2)} (r_{\rm m}) =\, a_{3} &\mathbf{\Psi}_{\rm{out}}^{(1)} (r_{\rm m}) + a_{4} \mathbf{\Psi}_{\rm{out}}^{(2)} (r_{\rm m}) \nonumber \\
     & + a_{5} \mathbf{\Psi}_{\rm{out}}^{(3)} (r_{\rm m})
\end{align}
%%%%%%%%%%%%%%%
The matching condition above represents four scalar equations, one for each component of $\mathbf{\Psi}$, for the five coefficients $a_{1}$, $a_{2}$, $a_{3}$, $a_{4}$, and $a_{5}$. Since the overall normalization of the perturbation is arbitrary, one coefficient may be fixed, and the remaining four coefficients are obtained by solving the resulting $4 \times 4$ linear system. This determines the relative weights of the inward- and outward-integrated basis solutions and therefore constructs, for the chosen trial frequency $\omega$, a global single interior solution that is regular at the center, satisfies the fluid-surface conditions, and is continuous at the matching radius.

Once this global interior solution has been constructed, the values of $H_{1}$ and $K$ at the outer stellar surface $R_{O}$ are obtained from it and used to initialize the exterior Zerilli--Regge--Wheeler construction described in Sec.~\ref{sec:2e}. This completes the evaluation of the continued-fraction function ${\cal F} (\omega)$ for the chosen trial frequency, which will generically not be zero, and we thus, hereafter, refer to as the residual. By repeating the same procedure over a range of trial frequencies, we obtain $|{\cal F} (\omega)|$ as a function of $\omega$. The oscillation frequencies are identified from the deep minima of this function, and the associated eigenfunctions are obtained from the corresponding global interior solutions. The numerical workflow is summarized schematically in Fig.~\ref{fig:numerical_flowchart}. In the present implementation, the scan is performed along the real-frequency axis; the imaginary part of the quasi-normal mode frequency, which determines the gravitational-wave damping time, is not extracted here.
%%%%%%%%%%%%%%%

%%%%%%%%%%%%%%%%%%%%%%%%%%%%%
\section{Application to mirror dark matter admixed neutron stars}
\label{sec:4}
%%%%%%%%%%%%%%%%%%%%%%%%%%%%%
%%%%%%%%%%%%%%%
%%%%%%%%%%%%%%%%%%%%%%%%%%%%%
\subsection{Residual spectra and mode identification}
%\subsection{Two-fluid residual spectra and mode classification}
\label{sec:4a}
%%%%%%%%%%%%%%%%%%%%%%%%%%%%%
In this section, we present the first application of the formalism developed above to gravitationally coupled two-fluid neutron stars admixed with mirror dark matter. In the present work, by ``mirror dark matter'' we mean an \textit{idealized} realization in which the ordinary and dark components are assigned the \textit{same} QHC21-BT equation of state~\cite{Kojo_2022, TOGASHI201778}, while being treated as dynamically distinct fluids coupled only through gravity. This choice is not intended to represent the full range of possible mirror-sector microphysics; rather, it allows us to isolate the effects of the two-fluid structure and gravitational coupling without simultaneously varying an independent dark-sector equation of state. The formalism developed here, however, is not restricted to this specific case and applies more generally to any two-component neutron star model in which the constituents interact only gravitationally and can be treated as dynamically independent fluids. The results presented below therefore serve both as a concrete demonstration of the numerical implementation and as a first characterization of the polar non-radial mode structure in gravitationally coupled two-fluid stars.

%%%%%%%%%%%%%%%%%%%%%%%%%%%%%
%  Figure 1
%%%%%%%%%%%%%%%%%%%%%%%%%%%%%
\begin{figure}[htbp]
    \centering
    \includegraphics[width=\columnwidth]{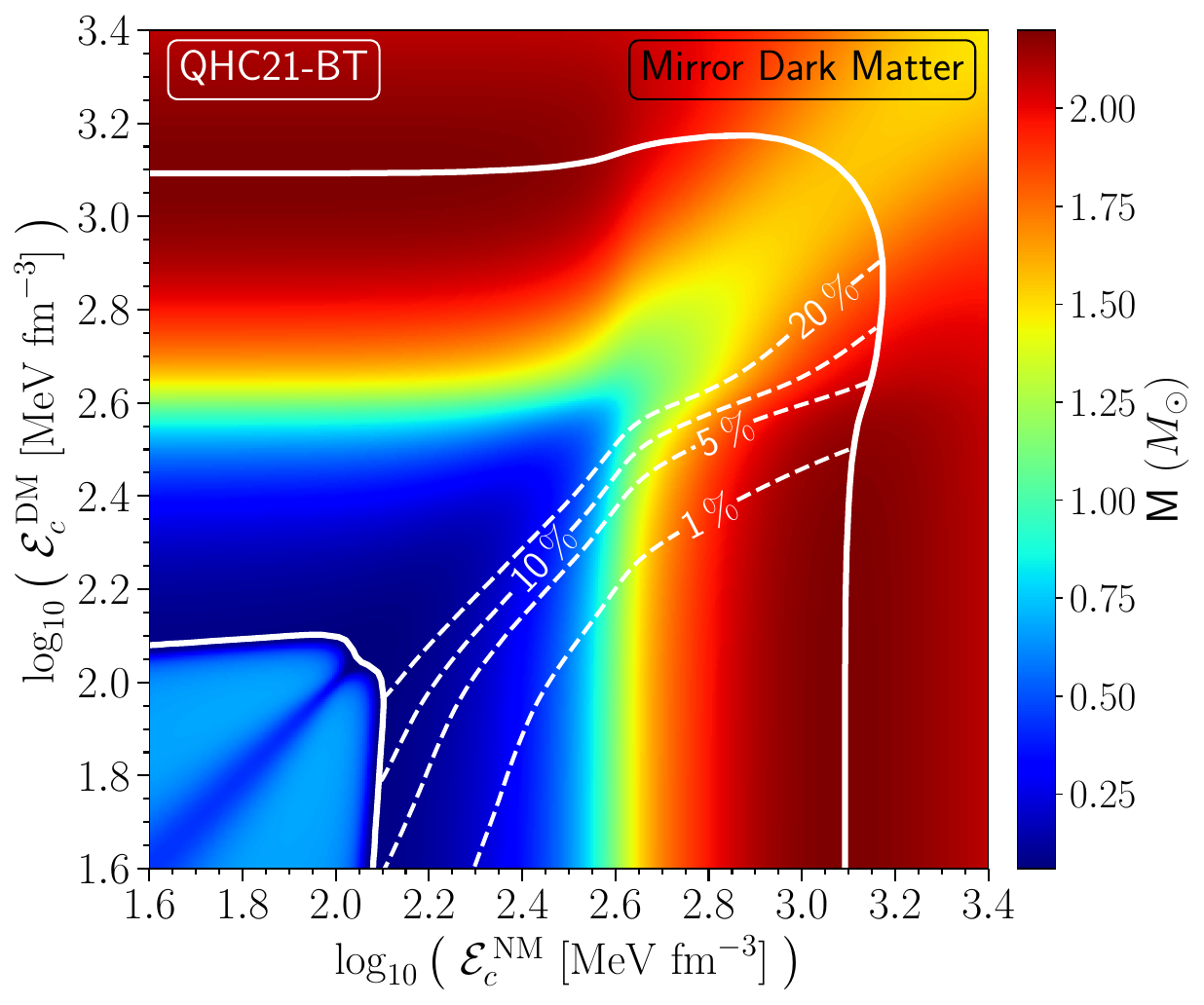}
    \caption{Gravitational mass contours of two-fluid neutron stars with mirror dark matter constructed using the QHC21-BT equation of state. The color map shows the total stellar mass $M$ (in units of $M_{\odot}$) in the plane spanned by the central energy densities of the nuclear matter (${\cal E}_{c}^{\rm{NM}}$) and dark matter (${\cal E}_{c}^{\rm{DM}}$) components. Solid white curves delineate the stability boundary of two-fluid equilibrium configurations; dynamically stable equilibria exist only within the region enclosed by these curves, while configurations outside this domain are unstable~\cite{Kumar:2025oyx}. Dashed white curves indicate contours of fixed dark matter mass fraction $\rm{F}_{\rm{DM}} \equiv (M_{\rm{DM}}/M) \times 100 = 1\%,\, 5\%,\, 10\%,$ and $20\%$. The mirror dark matter adopted here assigns the dark and nuclear components the same QHC21-BT equation of state.}
    \label{fig:mass_contour}
\end{figure}
%%%%%%%%%%%%%%%

Figure~\ref{fig:mass_contour} shows the equilibrium configuration space of mirror dark matter admixed two-fluid neutron stars constructed with the QHC21-BT equation of state in the plane of the central energy densities of the nuclear matter and dark matter components, $({\cal E}_{c}^{\rm{NM}},\,{\cal E}_{c}^{\rm{DM}})$. These equilibrium configurations are obtained by solving the two-fluid TOV equations, Eqs.~\eqref{eq:TOVphi}--\eqref{eq:TOVp}, for each pair of central densities. The color scale represents the total gravitational mass $M$ of the star, expressed in units of $M_{\odot}$. Owing to the mirror dark matter setup adopted here, where the nuclear matter and dark matter components are assigned the same equation of state, the mass contour structure is symmetric about the line ${\cal E}_{c}^{\rm{NM}}={\cal E}_{c}^{\rm{DM}}$. The solid white curves mark the stability boundary of the two-fluid equilibrium configurations, as determined from the radial oscillation analysis of Ref.~\cite{Kumar:2025oyx, Caballero:2024qtv}. Stable equilibrium configurations are confined within the region enclosed by these curves, whereas configurations outside this region are unstable. In particular, the stable region occupies a finite and nontrivial domain in the $({\cal E}_{c}^{\rm{NM}},\,{\cal E}_{c}^{\rm{DM}})$ plane, rather than a simple rectangular area, reflecting the genuinely two-dimensional nature of the stability problem in gravitationally coupled two-fluid stars. The slight rightward shift of the upper stability branch at intermediate ${\cal E}_{c}^{\rm{DM}}$ is another manifestation of this coupled response: the dark component modifies the mass distribution, pressure support, and gravitational potential, allowing stable models with somewhat larger ${\cal E}_{c}^{\rm{NM}}$ than in the nearly dark-matter-free limit; see also Ref.~\cite{Kumar:2025oyx}. The dashed white curves trace sequences of fixed dark matter mass fraction, $\rm{F}_{\rm{DM}} \equiv (M_{\rm{DM}}/M)\times 100$, where $M_{\rm{DM}}$ denotes the gravitational mass of the dark matter component and $M$ is the total stellar mass; the curves shown here correspond to $\rm{F_{DM}} = 1\%,\, 5\%,\, 10\%\,$, and $20\%$, and illustrate how configurations with the same global dark matter content are distributed across the two-dimensional central energy density plane. These fixed-$\rm{F}_{DM}$ sequences will be particularly useful in the following analysis, since they allow the oscillation spectra to be compared across equilibrium models with systematically varying dark matter content. 
%Observe that, as the dark matter central density is increased from very small values to moderate values, it changes the distribution of mass, pressure support, and gravitational potential in a nontrivial way. Because of this coupled response, the stability boundary can shift slightly so that configurations with somewhat higher ${\cal E}_{c}^{\rm{NM}}$ than the maximum for nearly DM-free limit remain stable when dark matter is included (see also~\cite{Kumar:2025oyx}).

%%%%%%%%%%%%%%%%%%%%%%%%%%%%%
%  Figure 2
%%%%%%%%%%%%%%%%%%%%%%%%%%%%%
\begin{figure*}[tbp]
%    \centering
    \includegraphics[width=\textwidth]{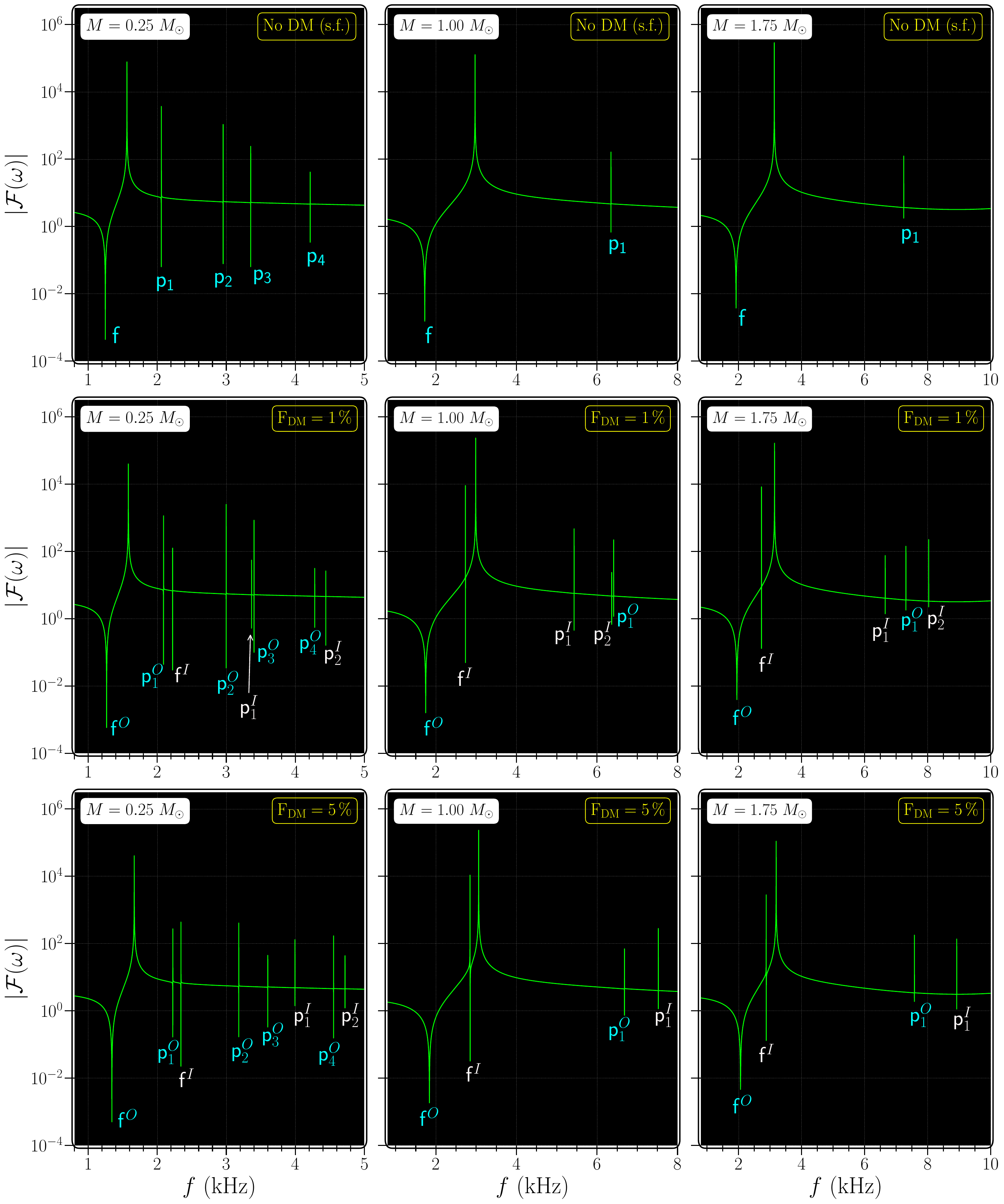}
    \caption{Boundary-condition residual (error function) $|{\mathcal F}(\omega)|$ as a function of the real trial frequency $f = \omega/2\pi$ for polar non-radial oscillations. The top row corresponds to single-fluid (s.f.) neutron stars constructed using the QHC21-BT equation of state, while the middle and bottom rows show gravitationally coupled two-fluid configurations with mirror dark matter based on the same microphysics, for dark matter mass fractions $\rm{F}_{\rm{DM}} = 1\%$ and $5\%$, respectively. Each panel shows the scanning result for a fixed total mass $M = 0.25\, M_{\odot}$ (left column), $M = 1.00\, M_{\odot}$ (middle column), and $M = 1.75\, M_{\odot}$ (right column). The eigenfrequencies are identified by the deep minima of $|{\mathcal F}(\omega)|$ along the real-frequency axis, which locate the real parts of weakly damped modes; the imaginary parts associated with gravitational-wave damping are not extracted in this scan. Labels indicate the dominant outer/nuclear- and inner/dark-fluid character of the fundamental and pressure modes, highlighting the hybridization of modes induced by the presence of a second gravitationally coupled fluid.}
    \label{fig:errfnc}
\end{figure*}
%%%%%%%%%%%%%%%

Figure~\ref{fig:errfnc} shows the boundary-condition residual, $|{\mathcal F}(\omega)|$, as a function of the real trial frequency $f=\omega/2\pi$ for representative single-fluid (s.f.) and mirror dark matter two-fluid neutron star models. The three columns correspond to fixed total masses $M=0.25\,M_{\odot}$, $1.00\,M_{\odot}$, and $1.75\,M_{\odot}$. The low-mass $0.25\, M_{\odot}$ model is included only as a diagnostic configuration: it makes several mode branches and their eigenfunction node structure visible within the frequency range shown, and is not intended to represent an observationally-typical neutron star. The rows compare the standard single-fluid configurations constructed with the QHC21-BT equation of state (top) with gravitationally coupled two-fluid configurations having dark matter mass fractions ${\rm F}_{\rm DM}=1\%$ (middle) and $5\%$ (bottom). In the single-fluid case, the residual is obtained from the usual polar perturbation problem built on the corresponding single-fluid TOV background, which is recovered in the present framework when the second fluid is absent. For each stellar model, the eigenfrequencies are identified by the deep minima of the residual, which represent the frequencies at which the interior solution can be matched consistently to a purely outgoing exterior spacetime solution. In the standard single-fluid case, the residual exhibits the expected sequence of minima associated with the usual $\mathsf{f}$-- and $\mathsf{p}$--modes. Once a second gravitationally coupled fluid is introduced, however, the structure of the spectrum becomes richer: additional minima appear, the original single-fluid sequence is supplemented by additional branches associated with the inner and outer fluid components, and the ordering of the pressure modes changes with both stellar mass and dark matter fraction. Even at the level of the residual function, the emergence of these extra deep minima already makes clear that the two-fluid system supports a substantially more intricate polar oscillation spectrum than its single-fluid counterpart. 

In principle, the quasi-normal mode frequencies of relativistic stars are complex, and may be written as $\omega = \omega_{R} + i\, \omega_{I}$, where the real part $\omega_{R}$ determines the oscillation frequency and the imaginary part $\omega_{I}$ encodes the damping due to gravitational-wave emission. In the present analysis, however, we restrict the scan in Fig.~\ref{fig:errfnc} to real trial frequencies. This is adequate for the purpose of identifying and comparing the mode spectrum, since for the fluid modes considered here, the imaginary part is typically much smaller than the real part ($|\omega_{I}| \ll |\omega_{R}|$), so that the locations of the deep minima of $|{\cal F}(\omega)|$ track the oscillation frequencies quite accurately, e.g., \cite{sotani2022}. The residual plots in Fig.~\ref{fig:errfnc} should therefore be understood as a practical and physically transparent way of resolving the real-frequency structure of the polar spectrum before turning to the detailed mode identification. Viewed in this way, the top row of Fig.~\ref{fig:errfnc} provides the standard single-fluid reference spectrum at the three representative masses considered here. For the lowest-mass model, $M = 0.25\, M_{\odot}$, one clearly resolves the fundamental mode together with a sequence of higher-order pressure modes (up to the $\mathsf{p}_{4}$--mode) within the scanned frequency range shown here, which extends up to 5 kHz. As the stellar mass is increased to $1.00\,M_{\odot}$ and $1.75\,M_{\odot}$, the scan is extended to 8 and 10 kHz, respectively, and the number and spacing of the visible minima change, but the residual still exhibits the familiar single-fluid polar mode structure, with a single $\mathsf{f}$--mode followed by the ordered sequence of $\mathsf{p}$--modes. Once mirror dark matter is introduced, however, the spectral structure changes both qualitatively and quantitatively. The two-fluid panels no longer display a single ordered branch of minima, but instead show multiple families associated with the outer nuclear matter fluid and the inner dark matter fluid, labeled here by the superscripts $O$ and $I$, respectively. For the configurations shown in Fig.~\ref{fig:errfnc}, the dark matter component remains the inner fluid, while the nuclear matter extends to the outer stellar surface, so this identification is unambiguous throughout the panels. The presence of these additional minima, however, does not by itself determine their fluid character, and a reliable assignment of the corresponding modes to the inner dark matter or outer nuclear matter component requires examination of the associated eigenfunctions.

\begin{figure*}[tbp]
%    \centering
    \includegraphics[width=\textwidth]{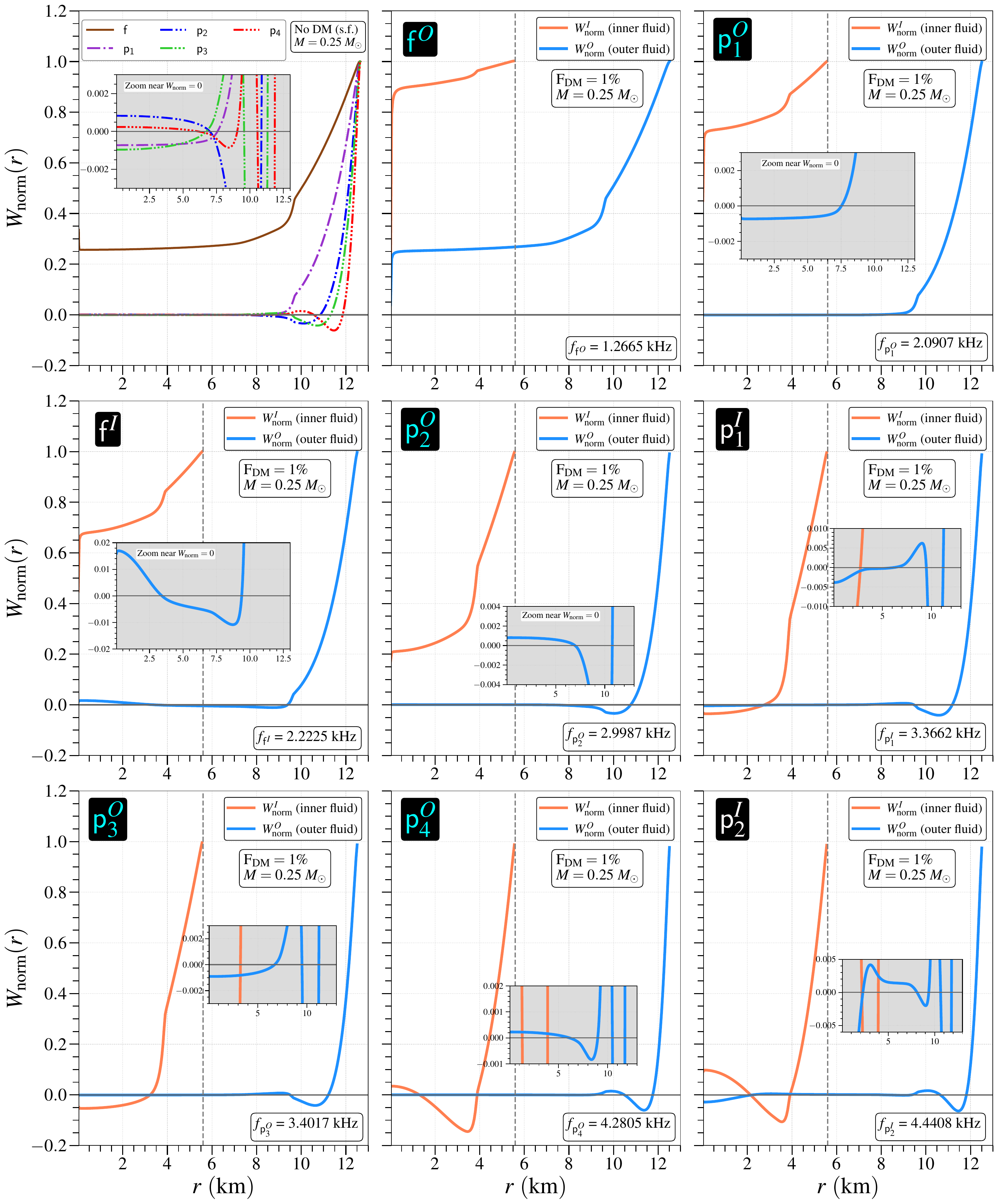}
    \caption{Normalized radial-displacement eigenfunctions for polar non-radial modes in the QHC21-BT model at fixed gravitational mass $M = 0.25\ M_{\odot}$. The top-left panel shows the reference single-fluid neutron star (no dark matter), with the normalized displacement $W_{\rm{norm}}$ for the $\mathsf{f}$ and $\mathsf{p_{1-4}}$ modes. The remaining eight panels correspond to a gravitationally coupled two-fluid configuration with mirror dark matter and dark matter mass fraction $\rm{F}_{\rm{DM}} = 1\%$, showing the normalized displacements of the inner (dark) and outer (nuclear) components, $W^{I}_{\rm{norm}}(r)$ and $W^{O}_{\rm{norm}}(r)$, for the identified mode families. Each panel is labeled by the mode assignment and the corresponding eigenfrequency (in kHz), as extracted from the deep minima of $|{\mathcal F}(\omega)|$: $\mathsf{f}^{O}$, $\mathsf{p}_{1}^{O}$, $\mathsf{f}^{I}$, $\mathsf{p}_{2}^{O}$, $\mathsf{p}_{1}^{I}$, $\mathsf{p}_{3}^{O}$, $\mathsf{p}_{4}^{O}$, $\mathsf{p}_{2}^{I}$. Vertical dashed lines mark the radius of the inner-fluid component. Insets zoom into the neighborhood of $W_{\rm{norm}} \simeq 0$ to make node structure and sign changes visible, enabling an unambiguous identification of which spectral feature corresponds to an outer- or inner-led mode.}
    \label{fig:eigenfnc_grid}
\end{figure*}

To make this assignment explicit, Fig.~\ref{fig:eigenfnc_grid} shows the normalized radial-displacement eigenfunctions\footnote{Throughout this section, the plotted displacement eigenfunctions are normalized by their value at the relevant surface: $W_{\rm{norm}} \equiv W/W(R)$ in the single-fluid case, $W^{I}_{\rm{norm}} \equiv W_{I}/W_{I}(R_{I})$ for the inner component, and $W^{O}_{\rm{norm}} \equiv W_{O}/W_{O}(R_{O})$ for the outer component.} associated with the minima identified in Fig.~\ref{fig:errfnc} for the $M = 0.25\, M_{\odot}$, $\rm{F}_{\rm{DM}} = 1\%$ configuration, together with the corresponding single-fluid reference spectrum at the same mass. The top-left panel of Fig.~\ref{fig:eigenfnc_grid} displays the normalized displacement eigenfunctions of the single-fluid star for the $\mathsf{f}$-- and $\mathsf{p}_{1-4}$--modes, while the remaining panels show the corresponding two-fluid eigenfunctions $W^{I}_{\rm{norm}}$ and $W^{O}_{\rm{norm}}$ for the eight deep minima appearing in the ${\rm{F}}_{\rm{DM}} = 1\%$ residual spectrum. In the two-fluid panels, the vertical dashed line marks the radius of the inner fluid, so that $W^{I}_{\rm{norm}}$ is defined only inside this boundary, whereas $W^{O}_{\rm{norm}}$ extends across the full stellar radius. In the single-fluid panel, the mode identification is straightforward: the $\mathsf{f}$--mode is the node-free solution, while the successive $\mathsf{p}$--modes are distinguished by the increasing number of radial nodes in the displacement eigenfunction, made particularly clear by the inset around $W_{\rm{norm}} \simeq 0$. This single-fluid pattern provides the reference against which the two-fluid eigenfunctions can be interpreted, particularly for identifying those modes whose outer-fluid displacement $W^{O}_{\rm{norm}}$ retains the characteristic node structure of the standard single-fluid spectrum.

%%%%%%%%%%%%%%%
%%%%%%%%%%%%%%%

For the two-fluid configurations, the outer-fluid-led modes ($O$--type) are identified by matching the node structure of $W^{O}_{\rm{norm}}$ to the standard single-fluid sequence, whereas the additional modes that do not follow this outer-fluid pattern are classified through the corresponding node structure of $W^{I}_{\rm{norm}}$ within the inner-fluid region. The lowest-frequency deep minimum is identified as the outer-fluid-led fundamental mode ($\mathsf{f}^{O}$--mode), since the corresponding outer-fluid eigenfunction $W^{O}_{\rm{norm}}$ remains node-free and follows the same qualitative pattern as the single-fluid fundamental mode. This identification is not based on node counting of $W^{O}_{\rm{norm}}$ alone, because the inner-fluid eigenfunction $W^{I}_{\rm{norm}}$ is also node-free within its domain; rather, the distinction becomes clear only after comparison with the eigenfunctions associated with the subsequent minima. The next minimum, whose associated eigenfunctions are shown in the upper-rightmost panel of Fig.~\ref{fig:eigenfnc_grid}, is identified as the first pressure mode of the $O$--led fluid ($\mathsf{p}^{O}_1$--mode), because the corresponding outer-fluid eigenfunction $W^{O}_{\rm{norm}}$ develops a single radial node, matching the characteristic $\mathsf{p}_1$ mode pattern of the single-fluid reference, even though the inner-fluid eigenfunction $W^{I}_{\rm{norm}}$ remains node-free within the inner-fluid domain in this case too. The following minimum, whose associated eigenfunctions are shown in the leftmost panel of the middle row of Fig.~\ref{fig:eigenfnc_grid}, is then identified as the inner-fluid-led fundamental mode ($\mathsf{f}^{I}$--mode): in this case, $W^{I}_{\rm{norm}}$ remains node-free within the inner-fluid region, while the accompanying outer-fluid eigenfunction $W^{O}_{\rm{norm}}$ no longer matches the characteristic pattern of the single-fluid $\mathsf{f}$-- or $\mathsf{p}$--mode sequence, and instead exhibits a distinctly distorted behavior near $W_{\rm{norm}} \simeq 0$, indicating that this mode does not belong to the outer-fluid-led branch. 

This also clarifies why the first two minima are not assigned to the inner-fluid-led ($I$--) branch: although $W^{I}_{\rm{norm}}$ remains node-free in both cases, the corresponding outer-fluid eigenfunctions $W^{O}_{\rm{norm}}$ still retain the standard single-fluid $\mathsf{f}$-- and $\mathsf{p}_{1}$--mode patterns, and are therefore naturally classified as $O$--led modes. Taken together, these two node-free minima therefore admit a natural interpretation as the fundamental modes of the outer and inner fluids, respectively: the lower-frequency one (at $1.2665$ kHz) preserves the standard outer-fluid $\mathsf{f}$--mode pattern and is identified as $\mathsf{f}^{O}$, whereas the higher-frequency one (at $2.2225$ kHz) is assigned to $\mathsf{f}^{I}$, since its outer-fluid eigenfunction no longer follows any of the standard single-fluid mode patterns.

The same comparison can be extended to the higher-frequency minima of the $M = 0.25\, M_{\odot}$, $\rm{F}_{\rm{DM}} = 1\%$ configuration shown in Fig.~\ref{fig:errfnc}, using the corresponding eigenfunctions plotted in the remaining panels of Fig.~\ref{fig:eigenfnc_grid}. The eigenmodes at $2.9987$, $3.4017$, and $4.2805$ kHz are identified as the $\mathsf{p}_{2}^{O}$--, $\mathsf{p}_{3}^{O}$-- and $\mathsf{p}_{4}^{O}$--modes, respectively, because the corresponding outer-fluid eigenfunctions $W^{O}_{\rm{norm}}$ reproduce the expected two-node, three-node, and four-node structure of the standard single-fluid $\mathsf{p}$-mode sequence. By contrast, the interleaved minima at $3.3662$ and $4.4408$ kHz are assigned to the $I$--led branch as the $\mathsf{p}_{1}^{I}$ and $\mathsf{p}_{2}^{I}$--modes, since the corresponding inner-fluid eigenfunctions $W^{I}_{\rm{norm}}$ exhibit one and two radial nodes, respectively, within the inner-fluid region, while the accompanying outer-fluid eigenfunctions $W^{O}_{\rm{norm}}$ no longer follow the characteristic pattern of the single-fluid $\mathsf{p}$--mode sequence, and instead exhibit a noticeably distorted behavior near $W_{\rm{norm}} \simeq 0$. The set of eight minima shown in Fig.~\ref{fig:eigenfnc_grid} therefore separates naturally into two interwoven mode sequences: one associated with the outer nuclear matter fluid, which preserves the ordering pattern of the standard single-fluid spectrum, and the other associated with the inner dark matter fluid, which appears through additional modes interleaved with the outer-fluid-led branches. The same identification procedure is used for the corresponding minima in the $\rm{F}_{\rm{DM}} = 1\%$ models at $M = 1.00\, M_{\odot}$ and $1.75\, M_{\odot}$, and likewise for the $\rm{F}_{\rm{DM}} = 5\%$ configurations shown in Fig.~\ref{fig:errfnc}.

%%%%%%%%%%%%%%%%%%%%%%%%%%%%%
\subsection{Mode frequencies and compactness universality}
\label{sec:4b}
%%%%%%%%%%%%%%%%%%%%%%%%%%%%%

\begin{figure}[htbp]
    \centering
    \includegraphics[width=\columnwidth]{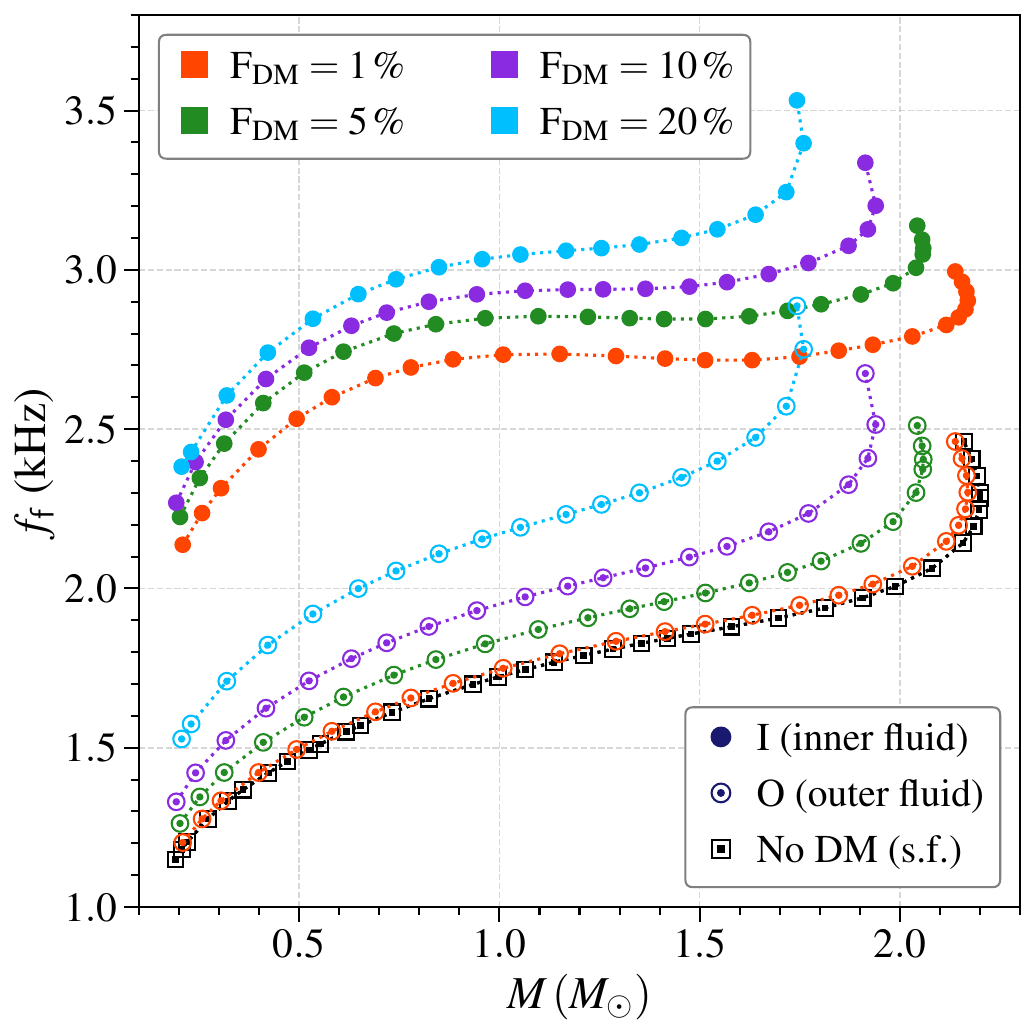}
    \caption{Fundamental mode ($\mathsf{f}$--mode) frequency $f_{\mathsf{f}}$ as a function of the gravitational mass $M$ for single-fluid neutron stars and gravitationally coupled two-fluid configurations with mirror dark matter constructed using the QHC21-BT equation of state. Filled colored circular markers show the $\mathsf{f}$--mode frequencies of the inner (dark matter) fluid led component, while ringed (bulls-eye) circular markers correspond to the outer (nuclear matter) fluid component. Different colors indicate dark matter mass fractions ${\rm F}_{\rm DM}=1\%,\ 5\%,\ 10\%$ and $20\%$, as shown in the legend. Black ringed square markers indicate the single-fluid (no dark matter) sequence. The two-fluid sequences exhibit a systematic separation of the fundamental mode into inner- and outer-fluid branches.}
    \label{fig:f-mode}
\end{figure}

\begin{figure}[htbp]
    \centering
    \includegraphics[width=\columnwidth]{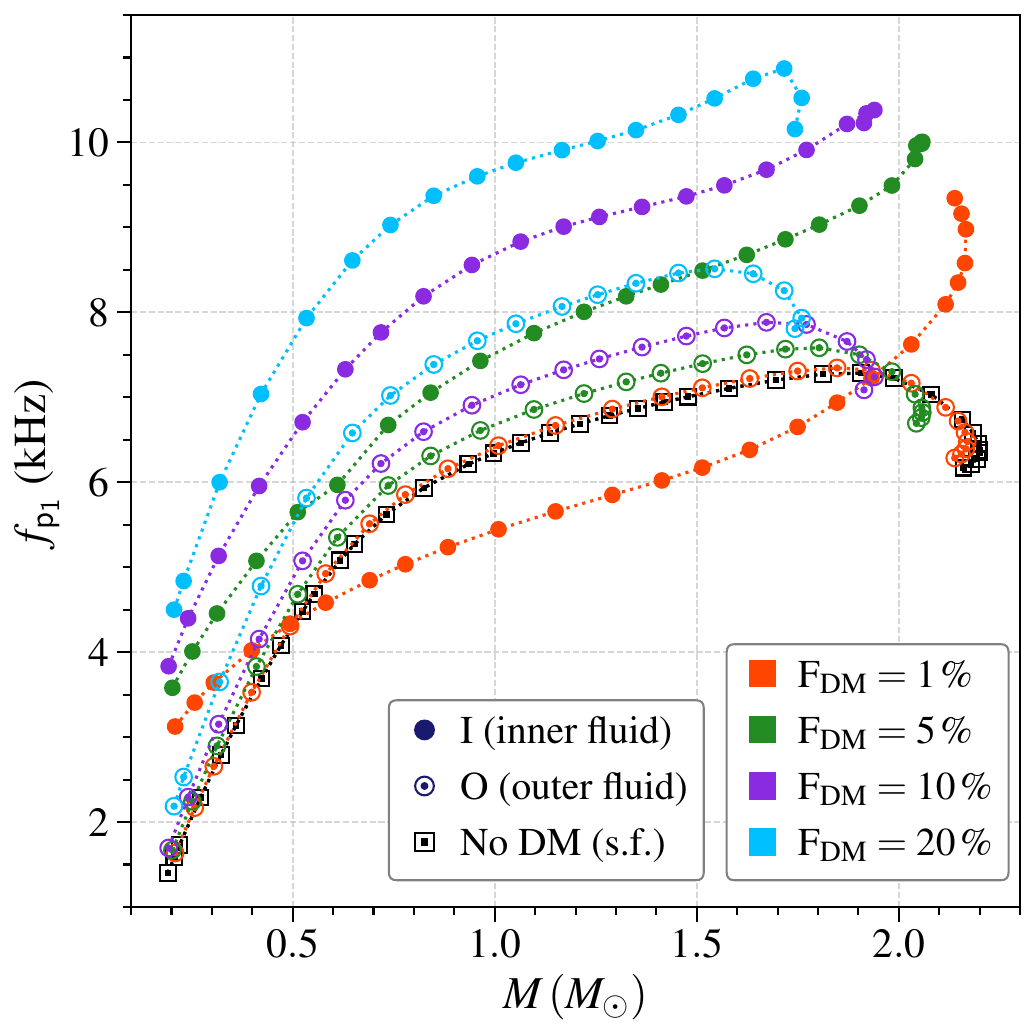}
    \caption{Same as Fig.~\ref{fig:f-mode}, but showing the first pressure mode ($\mathsf{p}_{1}$) frequency as a function of $M$ with the same marker and color conventions.}
    \label{fig:p1-mode}
\end{figure}

With the mode identification established, we now examine how the identified $\mathsf{f}$-- and $\mathsf{p}_{1}$--mode branches evolve with stellar mass and dark matter content. Figures~\ref{fig:f-mode} and \ref{fig:p1-mode} summarize, respectively, the mass dependence of the $\mathsf{f}$--mode and $\mathsf{p}_{1}$--mode frequencies for the single-fluid sequence and for gravitationally coupled two-fluid mirror dark matter configurations constructed with the QHC21-BT equation of state. In both figures, the black ringed square markers denote the standard single-fluid reference sequence, while the colored markers correspond to two-fluid configurations with fixed dark matter mass fractions $\rm{F}_{\rm{DM}} = 1\%,\ 5\%,\ 10\%$, and $20\%$. The filled circular markers represent the inner-fluid branch, which in the present models corresponds to the dark matter component, whereas the ringed circular markers denote the outer-fluid branch associated with the nuclear matter component. Taken together, these two figures provide a direct summary of how the standard single-fluid $\mathsf{f}$-- and $\mathsf{p}_{1}$--mode sequences are reorganized into separate inner- and outer-fluid branches once a second gravitationally coupled fluid is present, with the branch assignment made using the eigenfunction-based identification described above.
 
%%%%%%%%%%%%%%%
%%%%%%%%%%%%%%%

Figure~\ref{fig:f-mode} shows that, in the single-fluid case, the fundamental-mode frequency increases smoothly with stellar mass for neutron stars constructed with the QHC21-BT equation of state, thereby providing a useful reference against which the two-fluid trends can be assessed. A clear feature of Fig.~\ref{fig:f-mode} is that the outer-fluid branch, associated with the nuclear matter component, remains relatively closer to this single-fluid reference,  whereas the inner-fluid branch, associated with the dark matter component, is shifted to systematically higher $\mathsf{f}$--mode frequencies over the full mass range considered here. At fixed mass, the inner-fluid $\mathsf{f}$--mode is always found at a higher frequency than the corresponding outer-fluid mode, while the latter continues to track the behavior of the standard single-fluid sequence more closely. This trend is expected because the outer-fluid branch is associated with the nuclear matter component that still extends across the stellar radius, so that its displacement samples the full radial extent of the configuration and therefore continues to govern the global bulk oscillation of the star in a way that closely resembles the standard one-fluid case. By contrast, the inner-fluid branch is associated with oscillations of the more compact dark matter core. Since this motion is confined to a smaller radial region, it is characterized by a shorter effective length scale, and the corresponding restoring response operates over that shorter scale, leading naturally to a higher characteristic frequency. As the dark matter fraction increases, the equilibrium configuration itself is modified more strongly by the presence of the inner dark matter component, so that the common gravitational field and the associated restoring dynamics of the two-fluid system depart progressively farther from the single-fluid case. This is reflected in the progressively larger displacement of both two-fluid $\mathsf{f}$--mode branches away from the single-fluid reference as $\rm{F}_{\rm{DM}}$ increases and the two-fluid character of the star becomes more pronounced.

%%%%%%%%%%%%%%%
%%%%%%%%%%%%%%%

Figure~\ref{fig:p1-mode} depicts the $\mathsf{p}_{1}$--mode frequency as a function of gravitational mass for the same single-fluid and two-fluid mirror dark matter sequences considered in Fig.~\ref{fig:f-mode}. In contrast to the $\mathsf{f}$--mode case, the behavior of the $\mathsf{p}_{1}$--mode branches for these two-fluid stars is qualitatively different, especially for the inner-fluid sequence. The outer-fluid $\mathsf{p}_{1}$--branch, associated with the nuclear matter component, still follows the overall trend of the standard single-fluid sequence closely, but the inner-fluid branch behaves in a qualitatively different manner. For $\rm{F}_{\rm{DM}} = 1\%$, the inner-fluid $\mathsf{p}_{1}$--branch lies below both the outer-fluid branch and the single-fluid reference over a broad intermediate-mass range, before turning upward again toward the high-mass end. This more complicated behavior reflects the stronger sensitivity of pressure modes to where the compressional energy is localized between the extended nuclear layer and the inner dark matter region, so small changes in the two-fluid stratification can reorganize the $\mathsf{p}_{1}$ branch more strongly than the $\mathsf{f}$ branch. For larger dark matter fractions, by contrast, the inner-fluid $\mathsf{p}_{1}$--branch lies above both the outer-fluid branch and the single-fluid reference over the full mass range considered here, with its frequency generally increasing as $\rm{F}_{\rm{DM}}$ is raised from $5\%$ to $20\%$. Overall, the $\mathsf{p}_{1}$--mode shows a stronger sensitivity than the $\mathsf{f}$--mode to the detailed two-fluid structure of the star, rather than primarily to its overall bulk motion.

\begin{figure}[htbp]
    \centering
    \includegraphics[width=\columnwidth]{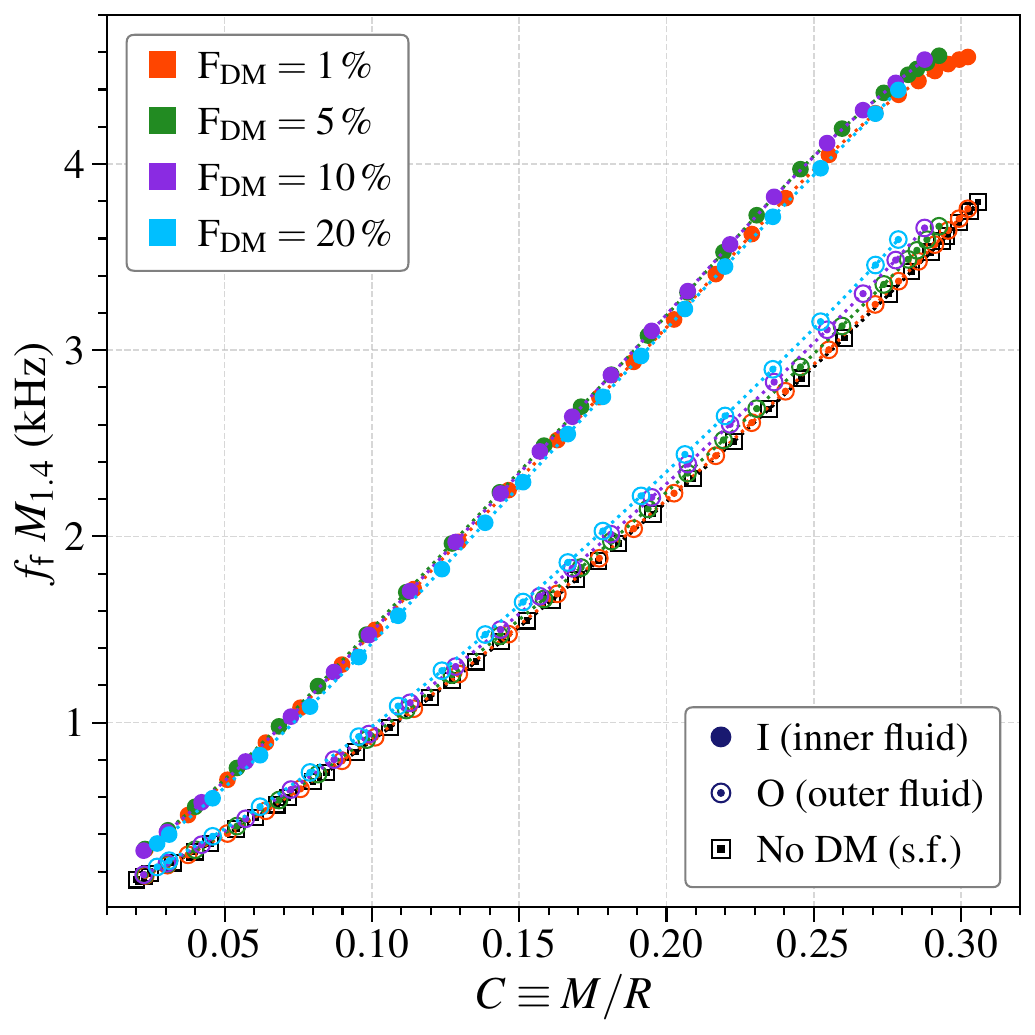}
    \caption{Mass-scaled fundamental-mode relation for mirror dark matter two-fluid stars constructed with the QHC21-BT equation of state. The $\mathsf{f}$--mode frequency is rescaled as $f_{\mathsf{f}}\,M_{1.4}$ with $M_{1.4}\equiv M/(1.4\,M_\odot)$, and plotted against the stellar compactness $C\equiv M/R$. Colored points correspond to gravitationally coupled two-fluid configurations with different colors indicating the dark matter mass fraction ${\rm F}_{\rm DM}=1\%,\ 5\%,\ 10\%$ and $20\%$ (see legend). For each $\rm F_{\rm DM}$ sequence, filled (solid) circular markers show the inner-fluid branch, while ringed circular markers show the outer-fluid branch, as indicated in the marker-style legend box. Black ringed square markers denote the standard single-fluid (no dark matter) sequence.}
    \label{fig:fMvsC}
\end{figure}

Figure~\ref{fig:fMvsC} presents the fundamental-mode results of Fig.~\ref{fig:f-mode} in the form commonly used to examine the compactness-based universal relation, by plotting the mass-scaled frequency $f_{\mathsf{f}}\,M_{1.4}$, with $M_{1.4}\equiv M/(1.4\,M_\odot)$, against the stellar compactness $C\equiv M/R$. For standard single-fluid neutron stars, this representation is known to yield a relation that is nearly insensitive to the equation of state between the $\mathsf{f}$--mode frequency and compactness~\cite{PhysRevD.103.123015}. Here, it provides a useful way to assess how the standard single-fluid behavior is altered in the two-fluid configurations, where a second gravitationally coupled fluid is present. The black ringed squares denote the single-fluid QHC21-BT sequence, while the colored points correspond to mirror dark matter two-fluid configurations at fixed ${\rm F}_{\rm DM}$, with filled and ringed circles representing the inner- and outer-fluid branches, respectively.

For the single-fluid sequence, the mass-scaled $\mathsf{f}$--mode frequency follows the expected monotonic increase with compactness, consistent with the standard compactness-based $\mathsf{f}$--mode relation for ordinary neutron stars.
%%%%%%%%%%%%%%%
%%%%%%%%%%%%%%%
The mirror dark matter two-fluid configurations, however, no longer collapse onto a single compactness sequence, but instead separate into two distinct branches: an outer-fluid branch that remains close to the single-fluid trend, and an inner-fluid branch that is shifted systematically to higher values of $f_{\mathsf{f}}\,M_{1.4}$ at the same compactness. This behavior is a direct consequence of the two-fluid character of the oscillation spectrum: even in the adopted mirror dark matter case, where both fluids are assigned the same equation of state, the two branches correspond to perturbations that are preferentially supported in different radial regions and are coupled through the common spacetime. As a result, the two-fluid configurations do not collapse onto the single-fluid compactness trend: configurations with comparable total compactness can exhibit different mass-scaled $\mathsf{f}$--mode frequencies depending on whether the mode belongs to inner- or outer-fluid-led branch. 

Another notable feature is the systematic evolution of this branch structure with increasing ${\rm F}_{\rm DM}$. In the total-compactness representation used in Fig.~\ref{fig:fMvsC}, this evolution is most clearly visible in the outer-fluid branch, which shifts progressively upward away from the single-fluid sequence as the dark component becomes more important. The corresponding change in the inner-fluid branch is less transparent in this projection, since the horizontal axis is the total stellar compactness $C \equiv M/R$, whereas the two mode branches are more naturally interpreted using the compactness scale of the fluid component that predominantly supports each branch. This can be seen more clearly in Fig.~\ref{fig:fM_vs_CI_CO} of Appendix~\ref{app:component_compactness}, where the same mass-scaled frequencies are plotted against the component-wise diagnostic compactnesses, $C_{I} \equiv M_{I}/R_{I}$ and $C_{O} \equiv M_{O}/R_{O}$, with $M_{I}$ and $M_{O}$ denoting the mass-function contributions of the inner and outer fluids, respectively. In that representation, increasing ${\rm F}_{\rm DM}$ drives both the inner- and outer-fluid sequences toward one another. This is the expected behavior as the mirror configuration approaches the symmetric case, ${\rm F}_{\rm DM} = 50\, \%$, where the two fluids have the same equation of state and contribute equally to the equilibrium stellar structure. In this limit, the system is invariant under interchange of the two components, so the distinction between inner- and outer-fluid-dominated oscillations no longer has a unique physical meaning. The fundamental mode is then more naturally understood as a collective oscillation of two identical gravitationally coupled fluids, rather than as two separately identifiable branches.

Taken together, these results show that the presence of a second fluid, even when it is assigned the same equation of state as the ordinary matter component and interacts only through gravity, qualitatively changes the non-radial oscillation spectrum of the star. In particular, the emergence of distinct inner- and outer-fluid-led branches and their departure from the standard mass-scaled $\mathsf{f}$--mode universal relation with compactness demonstrate that the usual single-fluid asteroseismic intuition cannot be carried over unchanged to gravitationally coupled two-fluid systems. The mirror dark matter setup adopted here therefore provides a particularly clean example of how additional fluid degrees of freedom can leave identifiable signatures in the mode spectrum, even in the absence of any non-gravitational coupling.

%%%%%%%%%%%%%%%%%%%%%%%%%%%%%
\section{Summary and outlook}
\label{sec:5}
%%%%%%%%%%%%%%%%%%%%%%%%%%%%%
%%%%%%%%%%%%%%%
In this work, we developed a fully relativistic framework for polar non-radial oscillations of gravitationally coupled two-fluid neutron stars in general relativity. The formulation assumes two independently conserved fluid components coupled only through the common spacetime, and thus extends relativistic stellar perturbation theory beyond the standard single-fluid description without introducing entrainment or direct microphysical interactions between the constituents. We derived the coupled linear perturbation equations for the metric and both fluid components, and completed the formulation by imposing regularity at the stellar center, boundary conditions at the individual fluid surfaces, and the matching conditions at the outer surface needed to connect the interior solution to the exterior vacuum spacetime and define the eigenvalue problem. Taken together, these ingredients provide a self-consistent general relativistic basis for studying non-radial oscillations in a broad class of gravitationally coupled multi-component compact stars.

To demonstrate the practical use of the formalism, we implemented the coupled perturbation system numerically and computed representative polar mode spectra for gravitationally coupled two-fluid stellar models. This implementation made it possible to solve the full two-fluid eigenvalue problem in the presence of spacetime coupling and to identify how the additional fluid degree of freedom reorganizes the structure of the non-radial mode spectrum. A central outcome of this implementation is that it provides a practical way to address mode identification in gravitationally coupled two-fluid stars, allowing the $\mathsf{f}$-- and $\mathsf{p}$--mode branches of the spectrum to be classified by their dominant inner- or outer-fluid character through the associated eigenfunctions and their node structure. In this way, the present framework goes beyond the formal derivation of the perturbation equations and establishes a workable strategy for computing and interpreting the polar oscillation spectrum of gravitationally coupled two-fluid stars in practice.

As a first application, we considered mirror dark matter admixed neutron stars constructed with the QHC21-BT equation of state. This choice isolates the effect of the two-fluid structure, since the nuclear and dark components obey the same equation of state, while remaining dynamically independent and coupled only through the spacetime geometry. The resulting non-radial spectrum differs qualitatively from the standard single-fluid one. The residual function develops additional minima associated with the extra fluid degree of freedom, and the corresponding eigenfunctions separate the spectrum into inner- and outer-fluid branches. For the fundamental mode, the outer-fluid branch remains closest to the single-fluid sequence, but is shifted systematically as the dark matter fraction increases, showing that the ordinary-matter-led oscillation is modified by the gravitational presence of the dark component. The inner-fluid branch appears at higher frequencies over the range of dark matter fractions considered, indicating that the second fluid introduces a distinct oscillation channel rather than merely perturbing the usual single-fluid mode. Consequently, the usual mass-scaled $\mathsf{f}$--mode relation with compactness does not remain a single universal sequence: configurations with similar total compactness can yield different values of $f_{\mathsf{f}}\,M_{1.4}$, depending on whether the mode belongs to the inner- or outer-fluid branch. This branch-dependent departure from the single-fluid relation provides one of the clearest signatures of the two-fluid character of the oscillation spectrum.

The framework developed here also opens several natural directions for future work. Although the present paper has focused on a mirror dark matter realization as a clean baseline application, the formalism itself is not restricted to that case and can be applied directly to more realistic two-fluid scenarios involving distinct equations of state and nontrivial dark-sector microphysics, including fermionic and self-interacting dark matter models. Such extensions will make it possible to explore how the detailed properties of the second fluid modify the non-radial spectrum, the organization of the mode branches, and the associated asteroseismic signatures in a quantitatively richer setting. In addition, while the present study has concentrated on representative polar fluid modes, the same general framework provides a basis for investigating other important sectors of the oscillation spectrum, including buoyancy-driven $\mathsf{g}$--modes and spacetime-dominated $\mathsf{w}$--modes, which are especially interesting in the present setting because both fluids couple through the same dynamical spacetime and can therefore influence the spacetime-led spectrum in a nontrivial way. More broadly, extending relativistic asteroseismology to genuinely multi-component compact stars may help clarify how additional internal degrees of freedom are encoded in oscillation spectra and, ultimately, how such effects could be probed through future gravitational-wave observations.
%%%%%%%%%%%%%%%
%%%%%%%%%%%%%%%%%%%%%%%%%%%%%%%%%%%%%%%%%%%%%%%%
\acknowledgments
%%%%%%%%%%%%%%%%%%%%%%%%%%%%%%%%%%%%%%%%%%%%%%%%
%%%%%%%%%%%%%%%
This work is supported in part by the Japan Society for the Promotion of Science (JSPS) KAKENHI grant Numbers 
JP23K20848  % Kiban(B) by Sotani
and JP24KF0090. % by Sotani & Kumar
D. A. C. is grateful for support from UIUC Graduate College and the Grainger College of Engineering, and from the Sloan Foundation.
N.Y. acknowledges the support of Simons Foundation International through grant SFI-MPS-BH00012593-01 and the National Science Foundation (NSF) through grant no. PHY-2512423.
%%N.Y. acknowledges support from the Simons Foundation through Award No. 896696, the Simons Foundation International through Award No. SFI-MPS-BH-00012593-01, the NSF through Grants No. PHY-2207650 and PHY-25-12423, and NASA through Grant No. 80NSSC22K0806.
%%%%%%%%%%%%%%%%%%%%%%%%%%%%%
%%%%%%%%%%%%%%%%%%%%%%%%%%%%%
\appendix
%%%%%%%%%%%%%%%%%%%%%%%%%%%%%
\section{Algebraic details of the perturbation equations}
\label{app:eqs}
%%%%%%%%%%%%%%%%%%%%%%%%%%%%%
%%%%%%%%%%%%%%%
In this appendix, we complement Sec.~\ref{sec:2c} by providing the intermediate algebraic steps leading to the derivation of Eqs.~\eqref{eq:dP}, \eqref{eq:contV} and \eqref{eq:contH}. These relations are used in Sec.~\ref{sec:2c} to close the first-order perturbation system for the variables $K$, $H_{1}$, $W_{X}$, and $\mathcal{P}_{X}$. Throughout this appendix, the harmonic time dependence and the summation over spherical-harmonic indices are suppressed, while the explicit $r^{l}$ scaling and the angular dependence through $Y^{lm}$ are retained.

\textbf{Derivation of the algebraic relation for $V_{X}$.} We first derive Eq.~\eqref{eq:contV} from the angular component of the perturbed Euler equation, Eq.~\eqref{eq:perteul}. We evaluate the angular component in index-lowered form. For polar perturbations, the $\theta$-component reduces to
%%%%%%%%%%%%%%%
\begin{equation}\label{eq:eulth}
    ({\cal E}_X+p_X)\,\delta a_\theta^X+\partial_\theta\, \delta p_X=0 \ ,
\end{equation}
%%%%%%%%%%%%%%%
where $\delta a_\theta^X$ is the Eulerian perturbation of the four-acceleration of fluid $X$. The corresponding Eulerian perturbation of the four-acceleration is
%%%%%%%%%%%%%%%
\begin{equation}\label{eq:accth}
    \delta a_\theta^X=-r^l\left(\tfrac12H_0-e^{-2\Phi}\omega^2V_X\right)\partial_\theta Y^{lm} \ .
\end{equation}
%%%%%%%%%%%%%%%
The pressure term (i.e., $\partial_\theta \delta p$) follows from the relation between Eulerian and Lagrangian perturbations,
%%%%%%%%%%%%%%%
\begin{align}\label{eq:dpxdth}
    \partial_\theta & \delta p_X = \partial_\theta\left[\Delta_X p_X-\xi^r_X\tfrac{d}{dr}p_X\right] \nonumber\\
    & = r^l\left[-e^{-\Phi}\mathcal P_X+e^{-\Lambda}\frac{{\cal E}_X+p_X}{r}\Phi'W_X\right]\partial_\theta Y^{lm} \ .
\end{align}
%%%%%%%%%%%%%%%
Substituting Eqs.~\eqref{eq:accth} and \eqref{eq:dpxdth} into Eq.~\eqref{eq:eulth}, and using the fact that the angular dependence factors out through $\partial_{\theta} Y^{lm}$, gives the algebraic relation for $V_{X}$,
%%%%%%%%%%%%%%%
\begin{align}
    ({\mathcal E}_X+p_X)\,\omega^2V_X =&\ e^{\Phi}\mathcal P_X-e^{2\Phi-\Lambda}\frac{{\mathcal E}_X+p_X}{r}\Phi'W_X \nonumber\\
    &\ +\frac12({\mathcal E}_X+p_X)e^{2\Phi}H_0 \ ,
\end{align}
%%%%%%%%%%%%%%%
which is Eq.~\eqref{eq:contV}.

\textbf{Derivation of the constraint for $H_0$.} We next derive the algebraic relation \eqref{eq:contH}, which determines $H_{0}$ in terms of the metric amplitudes $K$ and $H_{1}$, together with the fluid pressure variables ${\cal P}_X$. The derivation uses the $r\theta$ and $rr$ components of the linearized Einstein equations, since these components contain $dH_{0}/dr$ and $dK/dr$ and can be combined to obtain an independent expression for $dK/dr$. The $r\theta$ component of the perturbed Einstein tensor is
%%%%%%%%%%%%%%%
\begin{align}\label{eq:dGrth}
    \delta G_r^{\ \theta}=\frac{r^{l-2}}{2}\Bigg[&\frac{dH_0}{dr}-\frac{dK}{dr}+H_0\left(\frac lr+2\Phi'\right)+\frac lrK \nonumber\\
    &-re^{-2\Phi}\omega^2 H_1\Bigg] \partial_\theta Y_{lm}\ .
\end{align}
%%%%%%%%%%%%%%%
For the perfect-fluid stress-energy tensor considered here, there is no $r\theta$ component of the matter perturbation, so that $\delta T_{r}^{\ \theta} = 0$. The $r\theta$ component of the linearized Einstein equation therefore gives $\delta G^{\,\theta}_{r} = 0$, which provides one relation between $dH_{0}/dr$ and $dK/dr$.

The second relation is obtained from the $rr$ component of the linearized Einstein equations. The corresponding perturbation of the Einstein tensor is
%%%%%%%%%%%%%%%
\begin{widetext}
    \begin{align}\label{eq:dGrr}
        \delta G_r^{\ r}=r^l\Bigg\{&e^{-2\Lambda}\left[\left(\frac1r+\Phi'\right)\frac{dK}{dr}-\frac1r\frac{dH_0}{dr}\right] -K\left[\frac{(l-1)(l+2)}{2r^2}-e^{-2\Phi}\omega^2-e^{-2\Lambda}\frac lr\left(\frac1r+\Phi'\right)\right] \nonumber\\
        &+\frac{H_0}{r}\left[\frac{l(l+1)}{2r}- e^{-2\Lambda}\left(\frac{l+1}{r}+2\Phi'\right)\right] -2e^{-2\Lambda-2\Phi}\omega^2H_1\Bigg\}Y_{lm}
    \end{align}
\end{widetext}
%%%%%%%%%%%%%%%
The matter source in this component is given by the Eulerian pressure perturbation summed over all fluids,
%%%%%%%%%%%%%%%
\begin{align}\label{eq:dTrr}
    \delta T_r^{\ r} &= \sum_X\delta p_X \nonumber \\ 
    & = r^l\sum_X\left[-e^{-\Phi}\mathcal P_X+e^{-\Lambda}\frac{{\cal E}_X+p_X}{r}\Phi'W_X\right]Y^{lm} \ .
\end{align}
%%%%%%%%%%%%%%%
The $rr$ component of the linearized Einstein equations is therefore obtained by imposing $\delta G_{r}^{\ r} = 8\pi\, \delta T_{r}^{\ r}$. This gives the second relation involving $dH_{0}/dr$ and $dK/dr$, together with algebraic terms in $H_{0}$, $H_{1}$, $K$, $W_{X}$, and ${\mathcal P}_{X}$.

Together, the $r\theta$ and $rr$ components provide two linear algebraic equations for the derivative pair $dH_{0}/dr$ and $dK/dr$. Solving this pair for $dK/dr$, and using the background Einstein equations to simplify the result, gives an expression independent of Eq.~\eqref{eq:dK}:
%%%%%%%%%%%%%%%
\begin{widetext}
    \begin{align}
        \Phi'\frac{dK}{dr}=&K\left[e^{2\Lambda}\frac{(l-1)(l+2)}{2r^2}-e^{2\Lambda-2\Phi}\omega^2-\frac lr\Phi'\right]+e^{-2\Phi}\omega^2H_1-H_0\left[e^{2\Lambda}\frac{l(l+1)}{2r^2}-\frac{1}{r^2}\right] \nonumber\\
        &+8\pi e^{\Lambda}\sum_X\Phi'\frac{{\cal E}_X+p_X}{r}W_X-8\pi e^{2\Lambda-\Phi}\sum_X\mathcal P_X \  . \label{eq:dK2}
    \end{align}
\end{widetext}
%%%%%%%%%%%%%%%
Equation~\eqref{eq:dK2} is obtained solely from the $r\theta$ and $rr$ components of the linearized Einstein equations. It therefore provides an independent expression for $dK/dr$, whereas Eq.~\eqref{eq:dK} was obtained from the $tr$ component. Multiplying Eq.~\eqref{eq:dK} by $\Phi'$ and equating the result to Eq.~\eqref{eq:dK2} eliminates $dK/dr$. After collecting terms, one obtains the algebraic constraint equation for $H_{0}$, Eq.~\eqref{eq:contH}.

\textbf{Derivation of the differential equation for ${\mathcal P}_X$.} We finally derive Eq.~\eqref{eq:dP}, which gives the first-order evolution equation for the pressure variable ${\mathcal P}_X$. This equation follows from the radial component of the perturbed Euler equation, Eq.~\eqref{eq:perteul}. Evaluating the radial component in index-lowered form, the relevant terms reduce to
%%%%%%%%%%%%%%%
\begin{equation}\label{eq:Eulr}
    ({\cal E}_X+p_X)\delta a_r^X+(\delta{\cal E}_X+\delta p_X)a_r^X+\partial_r\delta p_X=0 \ .
\end{equation}
%%%%%%%%%%%%%%%
The Eulerian perturbation of the radial component of the four-acceleration is
%%%%%%%%%%%%%%%
\begin{align}
    \delta a_r^X&=r^l\Bigg[-\omega^2\frac{e^{\Lambda-2\Phi}}{r}W_X-\frac{l}{2r}H_0-\frac12\frac{dH_0}{dr} \nonumber\\
    &\quad\quad\quad-e^{-2\Phi}\omega^2rH_1\Bigg]Y^{lm}  \ .
\end{align}
%%%%%%%%%%%%%%%
The radial derivative of the Eulerian pressure perturbation follows from $\delta p_{X} = \Delta_{X}\,p_{X} - \xi_{X}^{r} (dp_{X}/dr)$, together with ${\mathcal P}_{X} = -e^{\Phi}\Delta_{X}\,p_{X}$ and the background Euler equation $dp_{X}/dr = -({\cal E}_{X}+p_{X})\Phi'$. This gives
%%%%%%%%%%%%%%%
\begin{align}
    \partial_r\delta p_X =&\ \frac{d}{dr}\Big[-r^l{\mathcal P}_Xe^{-\Phi} \nonumber \\
    &\ +e^{-\Lambda} r^{l-1}W_X({\cal E}_X+p_X)\Phi'\Big] Y^{lm}\ .
\end{align}
%%%%%%%%%%%%%%%
Substituting these expressions into Eq.~\eqref{eq:Eulr}, expanding the radial derivative of $\delta p_{X}$, and removing the common factor $r^lY^{lm}$, one obtains
%%%%%%%%%%%%%%%
\begin{widetext}\label{eq:dPApp}
    \begin{align}
        0=&-e^{-\Phi}\frac{d\mathcal P_X}{dr}+e^{-\Lambda}\frac{{\cal E}_X+p_X}{r}\Phi'\frac{dW_X}{dr}-\frac12({\cal E}_X+p_X)\frac{dH_0}{dr}-e^{-\Phi}\left[\frac lr+\frac{\Phi'}{c_X^2}\right]\mathcal P_X \nonumber\\
        &-({\cal E}_X+p_X)\left\{\frac{l}{2r}H_0+re^{-2\Phi}\omega^2H_1+\frac{e^{\Lambda}}{r}\left[e^{-2\Phi}\omega^2-e^{-2\Lambda}\Phi''+e^{-2\Lambda}\left(\Lambda'-\frac{l-1}{r}\right)\Phi'\right]W_X\right\}
    \end{align}
\end{widetext}
%%%%%%%%%%%%%%%
Here $c_{X}^{2} \equiv dp_{X}/d{\cal E}_{X}$ has been used to write $d{\mathcal E}_{X}/dr = (d{\mathcal E}_{X}/dp_{X})(dp_{X}/dr) = -({\cal E}_{X}+p_{X})\,\Phi'/c_{X}^{2}$. The intermediate equation still contains the derivatives $dW_{X}/dr$ and $dH_{0}/dr$. The first of these is eliminated using Eq.~\eqref{eq:dW}. To remove $dH_{0}/dr$, we use the $rr$ component of the linearized Einstein equations together with Eq.~\eqref{eq:dK}. Specifically, substituting Eq.~\eqref{eq:dK} into Eq.~\eqref{eq:dGrr} and imposing $\delta G_{r}^{\ r} = 8\pi\, \delta T_{r}^{\ r}$ gives
%%%%%%%%%%%%%%%
\begin{widetext}
\begin{align}\label{eq:dH0}
    \frac{dH_0}{dr} =&\ \left[e^{2\Lambda}\frac{l(l+1)}{2r}-\left(\frac lr+\Phi'\right)\right]H_0 + \left[\frac12 l(l+1)\left(\frac1r+\Phi'\right)-2e^{-2\Phi}r\omega^2\right]H_1 \nonumber\\
    &\ -r\left[e^{2\Lambda}\frac{(l-1)(l+2)}{2r^2}+\frac{1}{r^2}-(\Phi')^2-e^{2\Lambda-2\Phi}\omega^2\right]K+8\pi\sum_X\left[re^{2\Lambda-\Phi}\mathcal P_X+e^\Lambda \frac{\mathcal E_X+p_X}{r}W_X\right]\ .
\end{align}
\end{widetext}
%%%%%%%%%%%%%%%
Substituting Eqs.~\eqref{eq:dW} and \eqref{eq:dH0} into the intermediate radial Euler equation eliminates the remaining derivatives $dW_{X}/dr$ and $dH_{0}/dr$. After collecting terms, one obtains
%%%%%%%%%%%%%%%
\begin{widetext}
    \begin{align}
        \frac{d\mathcal P_X}{dr} = &\ -\frac lr \mathcal P_X-(\mathcal E_X+p_X)e^\Phi\left\{\Phi' \frac{l(l+1)}{r^2}V_X+\frac{l(l+1)}{4}\left(\frac1r+\Phi'\right)H_1+e^{2\Lambda}\frac{l(l+1)}{4r}H_0\right\} \nonumber\\
        &\ +(\mathcal E_X+p_X)e^\Phi\Bigg\{ rK\left[\frac{1}{2r^2}+e^{2\Lambda}\frac{(l+2)(l-1)}{4r^2}-\frac12e^{2\Lambda-2\Phi}\omega^2-\Phi'\left(\frac1r+\frac12\Phi'\right)\right] \nonumber \\
        &\ -\frac{W_X}{r}\left[e^{\Lambda-2\Phi}\omega^2-r^2\frac{d}{dr}\left(\frac{1}{r^2}e^{-\Lambda}\Phi'\right)\right]\Bigg\} 
        -4\pi (\mathcal E_X+p_X)\left[\sum_Ye^{\Lambda+\Phi}\frac{(\mathcal E_Y+p_Y)}{r}W_Y + re^{2\Lambda}\mathcal P_Y\right] \ .
    \end{align}
\end{widetext}
%%%%%%%%%%%%%%%
The last line contains the summed pressure perturbation $\sum_{Y} {\mathcal P}_{Y}$. This term is removed using the algebraic constraint \eqref{eq:contH}, which expresses the same pressure sum in terms of $H_{0}$, $H_{1}$, and $K$. Substituting Eq.~\eqref{eq:contH} into the expression above and simplifying gives the final first-order equation for ${\mathcal P}_{X}$, Eq.~\eqref{eq:dP}.

%%%%%%%%%%%%%%%
\begin{figure}[htbp]
    \centering
    \includegraphics[width=\columnwidth]{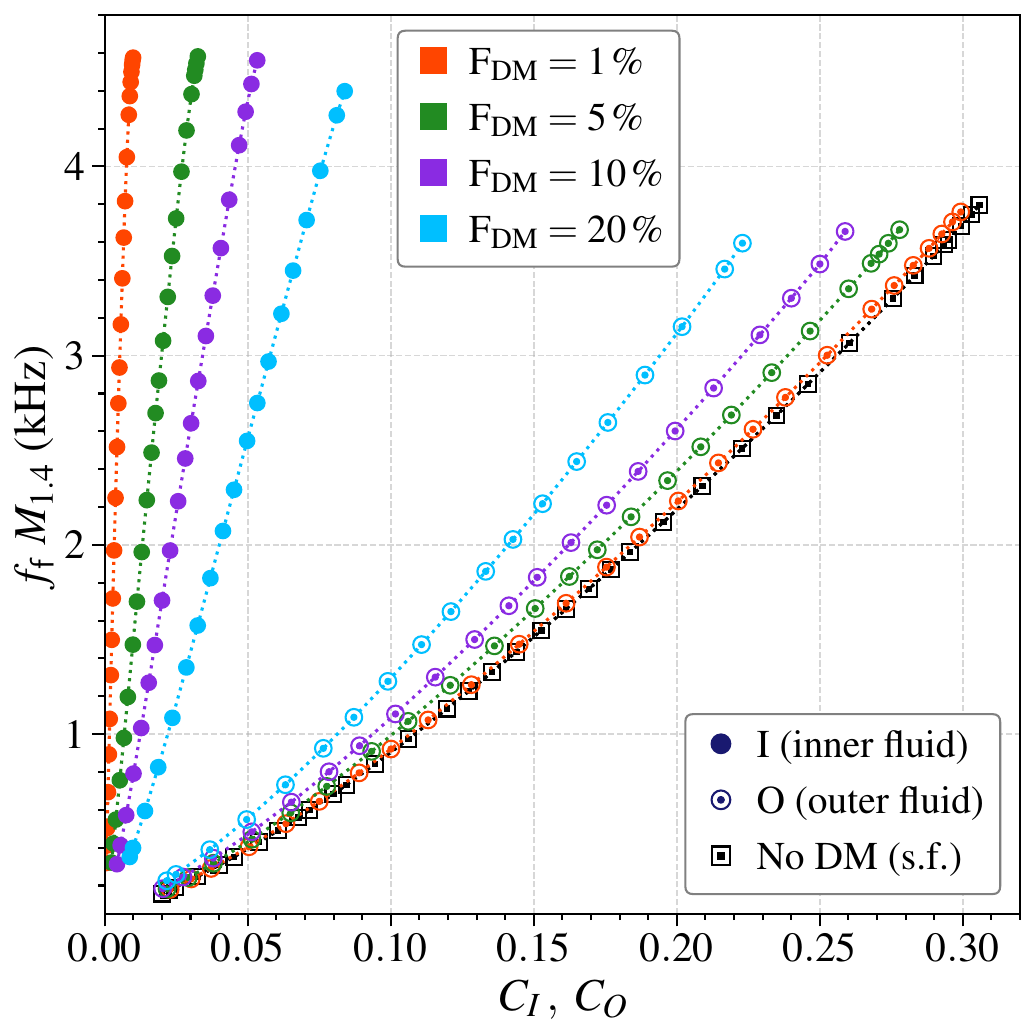}
    \caption{Mass-scaled fundamental-mode relation plotted against component-wise diagnostic compactnesses of the inner and outer fluid components for mirror dark matter two-fluid stars constructed with the QHC21-BT equation of state. The same quantity $f_{\mathsf{f}}\,M_{1.4}$ shown in Fig.~\ref{fig:fMvsC} is plotted here against the component-wise diagnostic compactness of each fluid component, $C_{I} \equiv M_{I}/R_{I}$ for the inner component and $C_{O} \equiv M_{O}/R_{O}$ for the outer component, where $M_{I}$ and $M_{O}$ denote the additive mass-function contributions of the corresponding fluids, as defined in the text, and $R_{I}$ and $R_{O}$ are the associated fluid surfaces. Filled circular markers show the inner-fluid branch, while ringed circular markers show the outer-fluid branch. Different colors correspond to fixed dark matter mass fractions ${\rm F}_{\rm DM}=1\%,\ 5\%,\ 10\%$ and $20\%$, and black ringed square markers denote the single-fluid QHC21-BT reference sequence. This representation makes clear that the evolution of the two branches with increasing $\rm F_{\rm{DM}}$ is more naturally interpreted in terms of the compactness scale of the fluid component that predominantly supports each mode branch.}
    \label{fig:fM_vs_CI_CO}
\end{figure}
%%%%%%%%%%%%%%%

%%%%%%%%%%%%%%%%%%%%%%%%%%%%%
\section{Mass-scaled $\mathsf{f}$--mode relation using individual fluid compactness}
\label{app:component_compactness}
%%%%%%%%%%%%%%%%%%%%%%%%%%%%%

The compactness-based relation shown in Fig.~\ref{fig:fMvsC} uses the total stellar compactness $C \equiv M/R$, where $M$ is the total gravitational mass and $R = R_{O}$ is the outer stellar radius. This choice is appropriate for testing the usual mass-scaled $\mathsf{f}$--mode universal relation with compactness, but it does not separately display the mass and length scales associated with each fluid component in a two-fluid configuration. Since the two branches in Fig.~\ref{fig:fMvsC} are identified through the inner- and outer-fluid perturbation variables, it is useful to examine the same mass-scaled frequencies using a component-wise compactness associated with the corresponding fluid sector. To construct this diagnostic, we define the additive mass-function contribution of each fluid as
%%%%%%%%%%%%%%%
\begin{align}
    M_{j} \equiv 4\pi \int_{0}^{R_{j}} {\cal E}_{j}(r)\ r^{2}\ dr\ , \qquad j=I,O\ ,
\end{align}
%%%%%%%%%%%%%%%
and introduce $C_{j} \equiv M_{j}/R_{j}$. These quantities should be understood as bookkeeping compactnesses associated with the individual fluid sectors, rather than as separately measurable gravitational compactnesses. In particular, the gravitational field at $R_{I}$ is determined by the total enclosed mass $M(R_{I})$, not only by $M_{I}$. Nevertheless, $C_{I}$ and $C_{O}$ provide a useful way to display how the two mode branches evolve with the characteristic mass and radius scales of the fluid component used to identify each branch.

Figure~\ref{fig:fM_vs_CI_CO} shows that the separation of the two branches in the mass-scaled $\mathsf{f}$--mode relation depends on the compactness variable used to represent the sequence. When plotted against the total compactness $C$, as in Fig.~\ref{fig:fMvsC}, increasing $\rm{F}_{\rm{DM}}$ produces a clear upward shift of the outer-fluid branch, while the corresponding evolution of the inner-fluid branch is less directly visible. The present representation separates these two component-wise scales by using $C_I$ for the inner-fluid branch and $C_O$ for the outer-fluid branch, thereby tracking the compactness scale associated with the fluid degree of freedom used to identify each branch. In this representation, increasing ${\rm F}_{\rm DM}$ moves the inner- and outer-fluid sequences toward one another, making the approach to the symmetric mirror configuration more transparent.

This comparison does not replace the total-compactness relation used in the main text, which remains the appropriate quantity for testing the standard single-fluid $f_{\mathsf f}M_{1.4}$--compactness universality. Rather, it provides a complementary diagnostic for two-fluid configurations by separating the mass-radius scales associated with the two fluid sectors. The result supports the interpretation given in Sec.~\ref{sec:4}: the split branches in Fig.~\ref{fig:fMvsC} arise from the two-fluid nature of the oscillation spectrum, while their evolution with increasing $\rm{F}_{\rm{DM}}$ reflects the gradual approach toward the symmetric mirror limit.

% References
%%%%%%%%%%%%%%%%%%%%%%%%%%%%%
\bibliographystyle{apsrev4-2}
\bibliography{main.bib} % Replace with your .bib file
\end{document}